\documentclass[twocolumn,amsmath,amssymb]{revtex4}

\newcommand{\vct}[1]{\mbox{\boldmath #1}}
\catcode`\@=11
\def\eqlt{\mathrel{\mathpalette\@vereq<}}  
\def\eqgt{\mathrel{\mathpalette\@vereq>}}  
\def\@vereq#1#2{\lower2.5pt\vbox{\baselineskip0pt \lineskip-.5pt
 \ialign{$\m@th#1\hfil##\hfil$\crcr#2\crcr{=}\crcr}}}

\newcommand{\simle}{\ \raise.3ex\hbox{$<$}\kern-0.8em\lower.7ex\hbox{$\sim$}\ }
\newcommand{\simge}{\ \raise.3ex\hbox{$>$}\kern-0.8em\lower.7ex\hbox{$\sim$}\ }

\usepackage{graphicx}
\begin{document}
\title {Orbital Physics in the Perovskite Ti Oxides}
\author{Masahito {\sc Mochizuki}$^1$ and Masatoshi {\sc Imada}$^{2,3}$}
\affiliation{$^1$Department of Physics, University of Tokyo, Hongo, Tokyo
113-0033, Japan\\
$^2$Institute for Solid State Physics, University of Tokyo, 
Kashiwanoha, Kashiwa, Chiba 277-8581, Japan\\
$^3$PRESTO, Japan Science and Technology Agency}
\date {today}

\begin{abstract}
Titanate compounds have been recognized as key materials for understanding the coupling of magnetism and orbitals in strongly correlated electron systems. In the perovskite Ti oxide $R$TiO$_3$ ($R=$trivalent rare-earth ions), which is a typical Mott-Hubbard insulator, the Ti $t_{2g}$ orbitals and spins in the $3d^1$ state couple each other through the strong electron correlations, resulting in a rich variety of orbital-spin phases. One way of controlling the coupling is to change the tiltings of the TiO$_6$ octahedra (namely the GdFeO$_3$-type distortion) by varying the $R$ ions, through which the relative ratio of the electron bandwidth to the Coulomb interaction is controlled.  With this control, these Mott insulators exhibit an antiferromagnetic-to-ferromagnetic (AFM-FM) phase transition, which has turned out to be a consequence of rich orbital physics in these materials. 
The origin and nature of orbital-spin structures of these Mott insulators have been intensively studied both experimentally and theoretically. When the Mott insulators are doped with carriers, the titanates show touchstone properties of the filling controlled Mott transition.
In this article, we first review the state of the art on the studies for understanding physics contained in the properties of the perovskite titanates. On the properties of the insulators, we focus on the following three topics: (1) the origin and nature of the ferromagnetism as well as the orbital ordering in the compounds with relatively small $R$ ions such as GdTiO$_3$ and YTiO$_3$, (2) the origin of the G-type antiferromagnetism and the orbital state in LaTiO$_3$, and (3) the orbital-spin structures in other AFM(G) compounds with relatively large $R$ ions ($R=$Ce, Pr, Nd and Sm). On the basis of these discussions, we discuss the whole phase diagram together with mechanisms of the magnetic phase transition. On the basis of the microscopic understanding of the orbital-spin states, we show that the Ti $t_{2g}$ degeneracy is inherently lifted in the titanates, which allows the single-band descriptions of the ground-state and the low-energy electronic structures as a good starting point. Our analyses indicate that these compounds offer good touch-stone materials described by the single-band Hubbard model on the cubic lattice. From this insight, we also reanalyze the hole-doped titanates $R_{1-x}A_x$TiO$_3$ ($A=$divalent alkaline-earth ions). Experimentally revealed filling-dependent and bandwidth-dependent properties and the critical behavior of the metal-insulator transitions are discussed in the light of theories based on the single-band Hubbard models.
\end{abstract}

\maketitle
\sloppy \maketitle
\section{Introduction}

After the discovery of high-$T_c$ superconductivity in the layered cuprate perovskites, strongly correlated electrons in the perovskite-type transition-metal oxides have attracted great interest~\cite{Imada98}. Extensive studies have revealed astonishing transport phenomena upon carrier doping in these compounds.

One of the important achievements of these studies is a discovery of the colossal magnetoresistance in the Mn perovskites. The manganites have the $e_g$ bands at the Fermi level similarly to the cuprate superconductors. However, it has turned out that the Mn $3d$ orbitals play crucial roles on the electronic properties of the manganites, which is in sharp contrast with the cuprate superconductors in which the orbital degree of freedom is fully quenched because of both $3d^9$ configuration of the mother compounds and large crystal-field splitting. While the strong electron correlations in the high-$T_c$ superconductors have been studied based on several single-band models including only the charge and spin degrees of freedom, those in the manganites have been studied by considering the coupling of not only charges and spins but also orbitals, which revived a  field of the condensed-matter physics, so called ``orbital physics"~\cite{Tokura00}.

Perovskite Ti oxide is also one of the important examples, and has attracted great interest. The titanates have the $t_{2g}$ bands at the Fermi level in contrast with the cuprates and manganites, and show very different properties from these two compounds in spite of the structural similarity. The perovskite titanate $R_{1-x}A_{x}$TiO$_3$ ($R$=trivalent rare-earth ions, $A$=divalent alkaline-earth ions) shows a transition from insulator to metal upon carrier doping achieved by increasing $x$. Thus, this system has been focused on as a typical example of the filling-control metal-insulator transition system~\cite{Tokura92,Tokura93a,Imada98}. The metallic phase proved to be well described by the Fermi liquid picture in marked contrast with the superconducting cuprates and the magnetoresistive manganites. As the system approaches the metal-insulator transition point from the metallic side, strong mass enhancement was experimentally found in the specific heat and the magnetic susceptibility. Anomalous filling-dependent behaviors near the transition point have been observed in various experiments, and have been discussed in connection to the nature of the Mott transition~\cite{Furukawa92,Imada93,Imada98}. The origin of the very different behavior of the titanates from the cuprates and manganites has been a subject of very intensive studies, and the titanates have been recognized as an important class of materials in clarifying the strong electron correlation effects.

In order to clarify the nature of the metallic phase and the transition, it is also important to clarify the electronic structures of the end Mott-Hubbard insulators $R$TiO$_3$. Similarly to the manganite case,  the Ti $3d$ orbitals have been recognized to be important for the electronic structures in $R$TiO$_3$ as well. In the Mott insulator, while the low-energy excitations of the charge degree of freedom are quenched by the insulating gap, the orbital and spin degrees of freedom survive, and the couplings and interplays of these two due to the strong electron correlations give rise to various electronic structures and puzzling phenomena. Coupling to lattice further enriches their interplays.

Because of a variety of orbital-spin phases and their phase transitions due to these couplings and interplays, the perovskite titanate $R$TiO$_3$ has been considered as a key class of materials for the understanding of the orbital physics. Because of the nominal three-fold degeneracy of $d^1$ state under the cubic crystal field on Ti ions in the insulating titanate, it is expected that the complete understanding of the Ti perovskites contributes in clarifying roles of the orbital degree of freedom, and couplings between magnetism and orbitals in the strongly correlated electron systems. In addition, the knowledge about the couplings is expected to open a possibility of controlling transport, optical and magnetic properties by various external control parameters like magnetic fields, electric fields, light, pressure etc. 

In this article, we first discuss recent progress of studies on the orbital-spin structures in the insulating titanates $R$TiO$_3$ and La$_{1-x}$Y$_x$TiO$_3$. We point out most of the available experimental results are accounted for by careful analyses of the orbital-spin structures and excitations. In particular, the realistic crystal field lifts the nominal $t_{2g}$ orbital degeneracy, providing a firm basis and justification of the description by a single-band Hubbard model. This indicates that the perovskite titanates indeed offer touch-stone materials for the single-band Hubbard model. From this justification, we revisit the bandwidth-control experiments for these compounds together with the theoretical analyses in detail and concludes that the single-band Mott transition is indeed realized in the titanates. In addition, we also reanalyze the experimental studies on the filling-control metal-insulator transition in the hole-doped systems by favorably comparing with theoretical results of single-band models. The limitation of the single-band description in real titanates is also discussed.  The rest of this article is organized as follows. We first introduce the experimentally obtained magnetic phase diagram for the insulating titanates, and mention some puzzles existing in this phase diagram. In Secs 3-6, by reviewing a number of attempts to solve these puzzles and the solutions provided by recent studies, we discuss the orbital-spin structures in the insulating titanates as well as the nature of their phase transition from a unified viewpoint. The bandwidth-control experiments and corresponding theoretical studies are also discussed in Sec. 6. In Sec. 7, we discuss the nature of the filling-control metal-insulator transition in the hole-doped systems based on both experimental and theoretical results. Section 8 is devoted to the concluding remarks.

\section{Magnetic phase diagram for the insulating titanates $R$TiO$_3$}

First, we briefly review some fundamental properties of the perovskite Ti oxides. In the end compound $R$TiO$_3$, Ti$^{3+}$ has a $t_{2g}^1$ configuration in which one of the threefold $t_{2g}$ orbitals is occupied by an electron.

\begin{figure}[tdp]
\caption{Crystal structure of $R$TiO$_3$ with the GdFeO$_3$-type distortion.}
\label{02fig01}
\end{figure}
The crystal structure of $R$TiO$_3$ is a pseudocubic perovskite with an orthorhombic distortion (the GdFeO$_3$-type distortion) in which the TiO$_6$ octahedra forming the perovskite lattice tilt alternatingly. Note that with this distortion, the cubic TiO$_6$ octahedra themselves are hardly distorted in contrast with the Jahn-Teller distortion. In this structure, the unit cell contains four TiO$_6$ octahedra (sites 1-4) as shown in Fig.~\ref{02fig01}. The magnitude of the distortion depends on the ionic radii of the $R$ ions. With a small ionic radius of the $R$ ion, the lattice structure is more distorted and the Ti-O-Ti bond angle is decreased more significantly from 180$^{\circ}$. For example, in LaTiO$_3$, the bond angle is 157$^{\circ}$ ($ab$-plane) and 156$^{\circ}$ ($c$-axis), but 144$^{\circ}$ ($ab$-plane) and 140$^{\circ}$ ($c$-axis) in YTiO$_3$~\cite{MacLean79}. The ionic radii of La and Y ions are 1.17 $\AA$ and 1.04 $\AA$, respectively. The magnitude of the distortion can also be controlled by using solid-solution systems. For example, by varying the Y concentration in La$_{1-x}$Y$_{x}$TiO$_3$, we can control the bond angle almost continuously from 157$^{\circ}$ ($x$=0) to 140$^{\circ}$ ($x$=1)~\cite{Okimoto95}. 

It is widely recognized that the tilting of the TiO$_6$ octahedra primarily controls the electron transfers between the neighboring Ti $t_{2g}$ orbitals mediated by the O $2p$ orbitals, thereby reduces the $t_{2g}$ bandwidth with decreasing bond angle. Therefore, the GdFeO$_3$-type distortion has been recognized as a bandwidth-control mechanism, and the variations of the Mott-Hubbard gap and electron correlations in the insulating titanates have been experimentally studied by controlling the bandwidth through this distortion~\cite{Okimoto95,Katsufuji95,Katsufuji97}. 

In addition to the bandwidth control, the tilting plays other important roles on the electronic structures. In fact, the GdFeO$_3$-type distortion controls the coupling and interplay of the orbital, spin and lattice degrees of freedom in the insulating titanates, and the system shows various orbital-spin phases and their phase transitions as we will discussed in the following.

\begin{figure}[tdp]
\caption{Experimentally obtained magnetic phase diagram for $R$TiO$_3$ (upper panel, Ref.~\cite{Katsufuji97,Greedan85}), and that for La$_{1-x}$Y$_x$TiO$_3$ (lower panel, Ref.~\cite{Okimoto95,Goral82}) from Ref.~\cite{Mochizuki04}. In thes phase diagrams, $T_{\rm N}$ and $T_{\rm C}$ are plotted as functions of the ionic radius of the $R$ ion and the unit cell volume, respectively. (The lines are guides for the eyes.) Note that the unit cell volume of La$_{1-x}$Y$_x$TiO$_3$ is approximately proportional to the Y concentration ($x$), and well characterizes the magnitude of the GdFeO$_3$-type distortion~\cite{Goral82}. Here, the open symbols and the closed symbols indicate $T_{\rm N}$ and $T_{\rm C}$, respectively. The values of the unit cell volume are deduced from the data obtained by the previous x-ray diffraction measurement (Ref.~\cite{Goral82}).}
\label{02fig02}
\end{figure}
A magnetic phase diagram in the plane of temperature and the magnitude of the GdFeO$_3$-type distortion was experimentally obtained for $R$TiO$_3$, which exhibits an antiferromagnetic (AFM)-to-ferromagnetic (FM) phase transition (see Fig.~\ref{02fig02} upper panel)~\cite{Katsufuji97,Greedan85}. LaTiO$_3$ with the smallest distortion shows a G-type AFM [AFM(G)] ground state, in which spins are aligned antiferromagnetically in all $x$, $y$ and $z$ directions as shown in the inset of the upper panel of Fig.~\ref{02fig02}. The magnetic moment is $\sim$0.57 $\mu_{\rm B}$~\cite{Cwik03}, which is reduced from the spin-1/2 moment of 1 $\mu_{\rm B}$. The N\'eel temperature ($T_{\rm N}$) is about 140-150 K. With increasing GdFeO$_3$-type distortion, $T_{\rm N}$ gradually decreases and is strongly depressed at SmTiO$_3$, subsequently a FM ordering appears. In the significantly distorted compounds such as GdTiO$_3$ and YTiO$_3$, a FM ground state accompanied by a large Jahn-Teller distortion is realized. A similar phase diagram was also obtained for La$_{1-x}$Y$_{x}$TiO$_3$ (see Fig.~\ref{02fig02} lower panel)~\cite{Okimoto95,Goral82}.

These phase diagrams provide us with several puzzling issues. First, strong suppressions of $T_{\rm N}$ and $T_{\rm C}$ around the AFM-FM phase transition point imply a continuous-type transition at $T=0$. At first sight, this is a puzzling feature, because we expect the first-order transition between completely different symmetry breakings at $T=0$, and $T_{\rm N}$ and $T_{\rm C}$ should remain nonzero at the transition point. Clarification of the mechanism of this continuous-type transition has long been a issue of intensive studies.

To clarify the nature and mechanism of this magnetic phase transition, it is essentially important to clarify the nature of the AFM(G) and FM phases. However, the origin of the AFM(G) ordering in the compounds with weak GdFeO$_3$-type distortion has also been a puzzle. In the orbitally degenerate systems, there is a strong tendency towards a FM state accompanied by an antiferro-orbital ordering~\cite{Imada98}, while $R$TiO$_3$ with $R=$La, Ce, Pr, Nd, Sm show actually AFM(G) order at low temperatures. Here, ``antiferro-orbital''order is defined as an order of orbital polarization with the staggered pattern in analogy with the staggered spin polarization in the AFM order. On the other hand, it turned out that the orbital state in the FM compounds with strong GdFeO$_3$-type distortion such as GdTiO$_3$ and YTiO$_3$ cannot be described by a simple antiferro-orbital ordering, which has also been a puzzle.

As discussed above, there exist a number of puzzles in the experimentally obtained phase diagram for the insulating titanates. It has been widely recognized that to fully understand the electron correlation effects in the perovskite compounds, these puzzles have to be explained by a convincing theory. In the following four sections (Secs.3-6), we discuss the orbital-spin structures in the insulating titanates as well as the nature of the phase diagram. In Sec. 3, we discuss the orbital ordering and the origin of the ferromagnetism in the FM compounds such as GdTiO$_3$ and YTiO$_3$, which are located in the region of strong GdFeO$_3$-type distortion. In Sec. 4, the AFM(G) ordering and the orbital state in LaTiO$_3$ with the smallest GdFeO$_3$-type distortion are discussed. In Sec. 5, the orbital-spin structures in other AFM(G) compounds ($R=$Ce, Pr, Nd and Sm) in the weakly distorted region are discussed. Finally, we discuss the whole nature of the magnetic phase diagram and the phase transition in Sec. 6.

\section{Orbital ordering and ferromagnetism in the compounds with strong GdFeO$_3$-type distortion}
In this section, we discuss the orbital-spin structures in the FM compounds with relatively small $R$ ions such as GdTiO$_3$ and YTiO$_3$. These compounds are located in the region of strong GdFeO$_3$-type distortion in the phase diagram of Fig.~\ref{02fig02}. Clarification of the origin and nature of the FM ordering is important for understanding the mechanism of puzzling second-order-like AFM-FM phase transition. Further, it is tightly connected with understanding the coupling between the magnetism and orbitals in the perovskite compounds. Hence, the issue has long attracted great interest, and a lot of studies have been done both theoretically and experimentally. We here overview the history of studies on this issue, and summarize recent progresses.

\begin{figure}[tdp]
\caption{Orbital ordering in YTiO$_3$. (From Akimitsu $et$ $al.$, Ref.~\cite{Akimitsu01})}
\label{03fig01}
\end{figure}
The emergence of the FM ordering in the titanates cannot be explained by a single-band model, and it has turned out indispensable to take the $3d$ orbital degeneracy into account to reproduce it. A model Hartree-Fock study based on the multiband $d$-$p$ model succeeded in reproducing the FM ordering in YTiO$_3$, and showed that this FM state is accompanied by an orbital ordering~\cite{Mizokawa96a,Mizokawa96b}. The first-principles band calculations based on the generalized gradient approximation (GGA) and the local spin-density approximation (LSDA) also succeeded in reproducing the FM state with the orbital ordering in YTiO$_3$~\cite{Sawada97,Sawada98}. In the predicted orbital ordering, the wavefunctions of the occupied orbitals at sites 1, 2, 3 and 4 are $c_1yz + c_2xy$, $c_1zx + c_2xy$, $c_1yz - c_2xy$ and $c_1zx - c_2xy$ (${c_1}^2+{c_2}^2=1$ and $c_1\sim \sqrt{0.5}$-$\sqrt{0.6}$, respectively (see also Fig.~\ref{03fig01}). This orbital ordering was actually observed in YTiO$_3$ by several experiments such as NMR (Refs.~\cite{Itoh97,Itoh99,Itoh01}) and polarized neutron scattering (Refs.~\cite{Ichikawa00,Akimitsu01}). A resonant x-ray scattering experiment (Ref.~\cite{Nakao02}) also succeeded in detecting this orbital ordering in YTiO$_3$ although we should mention that the mechanism of resonant x-ray scattering still has a controversy~\cite{Ishihara98a,Ishihara98b,Takahashi01}.

In the orbitally degenerate systems, there is a strong tendency towards FM state with antiferro-orbital ordering~\cite{Roth66,Kugel72,Kugel73,Inagaki73,Cyrot73,Cyrot75}. In the antiferro-orbital ordering, the neighboring occupied orbitals are orthogonal to each other. This orbital-spin configuration is stabilized by the spin-orbital exchanges due to both Hund's-rule coupling and electron transfers. Indeed, the occupied orbitals in YTiO$_3$ are nearly orthogonal in the $ab$-plane. However, the occupied orbitals along the $c$-axis are not necessarily orthogonal. In this sense, the emergence of the FM state in YTiO$_3$ cannot be understood straightforwardly as the ferromagnetism with antiferro-orbital ordering. The origin of the FM ordering was not clarified in the above theoretical studies. Moreover, the puzzle of the AFM-FM phase transition was also not resolved in these studies.

Recently, these issues were studied on the basis of an effective model in the insulating limit~\cite{Mochizuki00,Mochizuki01a}. This study showed that;
\begin{itemize}
\item The observed orbital ordering is stabilized by the hybridizations between the neighboring $e_g$ and $t_{2g}$ orbitals induced by the GdFeO$_3$-type distortion, which are prohibited without distortion.
\item The FM spin-exchange along the $c$-axis is realized by the super-exchange processes mediated by the $e_g$ orbitals due to the increased $t_{2g}$-$e_g$ hybridizations in the region of strong GdFeO$_3$-type distortion.
\item In the FM phase, the FM exchange along the $c$-axis becomes weaker with decreasing GdFeO$_3$-type distortion since the $t_{2g}$-$e_g$ hybridization is decreased. Hence, a two-dimensional anisotropy appears near the AFM-FM phase boundary, which results in the decrease of $T_{\rm C}$ as the system approaches the AFM-FM phase boundary.
\end{itemize}

Let us discuss some details of this study in the following: First, we briefly refer to the model used. The starting point is the multiband $d$-$p$ Hamiltonian which includes full degeneracies of the Ti $3d$ and O $2p$ orbitals. The Hamiltonian is given by
\begin{equation}
 H^{dp}=H_{d0}+H_{p}+H_{tdp}+H_{tpp}+H_{\rm on-site},
\label{dphamlt}
\end{equation}
The first and second terms denote the bare Ti $3d$ and the O $2p$ levels, respectively. The third and fourth terms represent the $d$-$p$ and $p$-$p$ transfers, respectively, which are determined by using the Slater-Koster parameters~\cite{Slater54,Harrison89}. The values of the Slater-Koster parameters were estimated in the cluster-model analysis of valence-band and transition-metal $2p$ core-level photoemission spectra~\cite{Saitoh95,Bocquet96} and the first-principles band calculation~\cite{Mahadevan96}. The effects of the lattice structure along with the lattice distortions are considered by modifying the transfer integrals. The fifth term expresses the on-site Coulomb interactions, which consists of the following four contributions;
\begin{equation}
 H_{\rm on-site}=H_{u}+H_{u'}+H_{j}+H_{j'}.
\label{onsiteClmb}
\end{equation}
Here, $H_{u}$ and $H_{u'}$ denote the intra- and inter-orbital Coulomb interactions, respectively, and $H_{j}$ and $H_{j'}$ denote the exchange interactions. The term $H_{j}$ is the origin of the Hund's-rule coupling which favors the spin alignment in the same direction on the same atoms. The term $H_{j'}$ gives the $\uparrow\downarrow$-pair hopping between the $3d$ orbitals on the same Ti atom.
These interactions are expressed by using Kanamori parameters, $u$, $u'$, $j$ and $j'$, which satisfy the following relations; $u=u'+2j$ and $j=j'$~\cite{Kanamori63,Brandow77}. Because of these relations, the Kanamori parametrization retains the rotational invariance. The values of these parameters were also estimated by the cluster-model analysis of the photoemission spectra~\cite{Saitoh95,Bocquet96}. By integrating out the O $2p$ orbitals in the path-integral formalism, we arrive at the multiband Hubbard model, which includes only Ti $3d$ orbitals and the transfer integrals between the Ti $3d$ orbitals mediated by the O $2p$ orbitals. By applying the second-order perturbational expansion with respect to the electron transfers between the Ti $3d$ orbitals to thus obtained multiband Hubbard Hamiltonian, we derive an effective spin-pseudospin Hamiltonian $H_{s\tau}$ on the subspace of states only with singly-occupied $t_{2g}$ orbitals at each Ti site in the limit of the strong Coulomb repulsion. In this effective Hamiltonian, only spin and orbital degrees of freedom are incorporated, and the threefold $t_{2g}$ orbitals are expressed by using the pseudospin-1 operator $\tau$. Further adding a crystal field term $H_{\rm cry.}$, we obtain the following effective spin-pseudospin Hamiltonian;
\begin{equation}
 H_{\rm eff.}=H_{\rm cry.}+H_{s\tau}.
\label{onsiteClmb}
\end{equation}

\begin{figure}[tdp]
\caption{$d$-type Jahn-Teller distortion.}
\label{03fig02}
\end{figure}
\begin{figure}[tdp]
\caption{Level splitting of the $3d$ orbitals in the cubic crystal field and that in the $d$-type Jahn-Teller distortion. With the $d$-type Jahn-Teller distortion, the threefold degenerate $t_{2g}$ level splits into twofold degenerate lower levels and a nondegenerate higher level at each Ti site. The twofold degenerate $e_g$ level also splits . The ways of the level splitting are different between sites 1, 3 and sites 2, 4.}
\label{03fig03}
\end{figure}
In order to study the FM phase in the region of strong GdFeO$_3$-type distortion, we have to take into account the effect of $d$-type Jahn-Teller distortion, which was experimentally observed in the FM compounds such as GdTiO$_3$ and YTiO$_3$. With the $d$-type Jahn-Teller distortion, the TiO$_6$ octahedra are elongated along the $y$ direction at sites 1 and 3, while at sites 2 and 4, they are elongated along the $x$ direction (see Fig.~\ref{03fig02}). The Ti $3d$ orbitals directed along the elongated bonds are lowered in energy as compared with the orbitals along the shorter bonds because of the reduction of the repulsive potential from the surrounding O ions. As a result, with the $d$-type Jahn-Teller distortion, the $xy$ and $yz$ orbitals are lowered at sites 1 and 3, and the $xy$ and $zx$ orbitals are lowered at sites 2 and 4 as shown in Fig.~\ref{03fig03}. The effect of the Jahn-Teller distortion can be treated in the multiband $d$-$p$ Hamiltonian by tuning the Slater-Koster parameters for the longer and shorter Ti-O bonds because it is empirically known that the magnitude of the transfer integral is proportional to $d^{-3.5}$ where $d$ being the bond length~\cite{Harrison89}.

\begin{figure}[tdp]
\caption{Total energies of the AFM(A), FM and AFM(G) states near the AFM(A)-FM phase boundary are plotted as functions of the Ti-O-Ti bond angle, which are calculated by using the effective spin-pseudospin Hamiltonian in the insulating limit (see text). Inset: energy difference between the AFM(A) and FM solutions. (From Mochizuki and Imada, Ref.~\cite{Mochizuki01a})}
\label{03fig04}
\end{figure}
Total energies of various orbital-spin configurations were calculated by solving the self-consistent mean-field equations. The result showed that in the region of strong GdFeO$_3$-type distortion, there exist AFM(A) and FM solutions accompanied by a certain type of orbital ordering as stable saddle-point solutions, and the AFM(A)-to-FM phase transition occurs with increasing GdFeO$_3$-type distortion as shown in the inset of Fig.~\ref{03fig04}. Note that the AFM(G) solution has much higher energies, and is unstable relative to the AFM(A) and FM solutions in this region.

As for the orbital state, the orbital ordering realized in the AFM(A) phase and that in the FM phase are similar to each other. We can specify the orbital states by using angles $\theta_{\rm AFM(A)}$ and $\theta_{\rm FM}$ as,
\begin{eqnarray}
&{\rm site}& 1;\quad\cos{\theta_x}|xy\rangle+\sin{\theta_x}|yz\rangle,
\nonumber \\
&{\rm site}& 2;\quad\cos{\theta_x}|xy\rangle+\sin{\theta_x}|zx\rangle,
\nonumber \\
&{\rm site}& 3;\quad-\cos{\theta_x}|xy\rangle+\sin{\theta_x}|yz\rangle,
\nonumber \\
&{\rm site}& 4;\quad-\cos{\theta_x}|xy\rangle+\sin{\theta_x}|zx\rangle,
\end{eqnarray}
where $x=$AFM(A), FM. Both $\theta_{\rm AFM(A)}$ and $\theta_{\rm FM}$ take $\sim45^{\circ}$, and the difference between them is very small. Namely, the orbital ordering in which $\frac{1}{\sqrt{2}}(xy+yz)$, $\frac{1}{\sqrt{2}}(xy+zx)$, $\frac{1}{\sqrt{2}}(-xy+yz)$ and $\frac{1}{\sqrt{2}}(-xy+zx)$ are occupied at sites 1, 2, 3, and 4, respectively, are realized in both AFM(A) and FM phases, and this orbital structure hardly changes through the AFM(A)-FM phase transition. We refer to this orbital ordering as ($yz$,$zx$,$yz$,$zx$)-type orbital ordering.

For the stabilization of this orbital ordering, the $e_g$ orbital degrees of freedom are indispensable. When the GdFeO$_3$-type distortion is absent, the hybridizations between the neighboring $t_{2g}$ and $e_g$ orbitals are prohibited by symmetry. As the GdFeO$_3$-type distortion increases, the symmetry restriction becomes to be relaxed, and the $t_{2g}$-$e_g$ hybridizations are induced. Since the hybridizations between the neighboring $e_g$ and O $2p$ orbitals have a $\sigma$-bonding character and are large, the amplitudes rapidly increase by the distortion. Calculation of the spin-independent perturbational energy showed that the ($yz$,$zx$,$yz$,$zx$)-type orbital ordering is actually stabilized by the $t_{2g}$-$e_g$ hybridizations induced by the GdFeO$_3$-type distortion. Indeed, it was shown that a model without the $e_g$ orbitals cannot reproduce this orbital ordering.

This orbital ordering was actually observed in YTiO$_3$ by several experiments (Refs.~\cite{Itoh97,Itoh99,Itoh01,Ichikawa00,Akimitsu01,Nakao02}), indicating that this FM solution corresponds to nothing but the FM phase in YTiO$_3$ whose Ti-O-Ti bond angle is $\sim 140^{\circ}$. On the other hand, the AFM(A) phase has not been observed in the actual compounds so far. However, the nature of this FM phase is well characterized through the clarification of the AFM(A)-FM transition. Thus, we discuss the mechanism and nature of this AFM(A)-FM phase transition in the following.

\begin{figure}[tdp]
\caption{Second-order perturbational energy gains strongly depend on the spin configuration. Substantial transfer processes along the $c$-axis for the energy gains are illustrated for both parallel and antiparallel spin configurations. The angles $\theta_{\rm AFM(A)}$ and $\theta_{\rm FM2}$ are fixed at 45$^{\circ}$. The cross symbol $\times$ represents the forbidden transfer process. (From Mochizuki and Imada, Ref.~\cite{Mochizuki01a})}
\label{03fig05}
\end{figure}
The AFM(A)-to-FM phase transition is identified as a transition where the sign of the spin-exchange constant along the $c$-axis changes from positive to negative while that in the $ab$-plane is always negative. The constant FM coupling in the $ab$-plane under the ($yz$,$zx$,$yz$,$zx$)-type orbital ordering can be easily understood. In the $ab$-plane, the neighboring orbitals are approximately orthogonal to each other. Hence, the FM spin configuration is favored through the Hund's-rule coupling. However, the emergence of the FM phase is not understood straightforwardly since the neighboring orbitals along the $c$-axis are not necessarily orthogonal.

In fact, the hybridizations of the $t_{2g}$ and $e_g$ orbitals again play an important role on the emergence of the FM phase. When the orbital state is strongly stabilized independently of the spin structure, the spin-exchange interaction is determined by the perturbational energy gain due to the electron transfers. In order to examine the spin-exchange interaction along the $c$-axis, let us consider the perturbational processes with respect to the electron transfers between the $3d$ orbitals on sites 1 and 3, which are neighboring along the $c$-axis. We fix the angles $\theta_{\rm AFM(A)}$ and $\theta_{\rm FM}$ at 45$^{\circ}$. Namely, we assume that an electron occupies $\frac{1}{\sqrt{2}}(xy+yz)$, $\frac{1}{\sqrt{2}}(-xy+yz)$ at sites 1 and 3, respectively.

In Fig.~\ref{03fig05}, we display the main perturbational processes for both parallel and antiparallel spin configurations. When the spins are parallel, the transfer process 2) is forbidden because of the Pauli principle. Therefore, the antiparallel spin configuration gains an energetical advantage of $t_2^2/u$ relative to the parallel spin configuration. On the other hand, in the transfer processes 1) and 3), the energies of the intermediate states are lowered by Hund's-rule coupling $j$. Therefore, the parallel spin configuration gains an advantage of,
\begin{eqnarray}
 \left(\frac{t_1^2}{u'-j}+\frac{t_3^2}{(u'-j+\Delta_{e_g})}\right)
-\left(\frac{t_1^2}{u'}+\frac{t_3^2}{(u'+\Delta_{e_g})}\right) \nonumber\\
 \sim \frac{t_1^2j}{{u'}^2}+\frac{t_3^2j}{(u'+\Delta_{e_g})^2},
\end{eqnarray}
relative to the antiparallel configuration. Here, $\Delta_{e_g}$ is the energy difference between the $t_{2g}$ and $e_g$ levels.

Then, the spin configuration along the $c$-axis is determined by the competition between these two energies, $t_2^2/u$ and $t_1^2j/{u'}^2 + t_3^2j/(u'+\Delta_{e_g})^2$, which favor the FM and AFM spin configurations, respectively, and the spin-exchange constant is proportional to the difference of these two energies. As the GdFeO$_3$-type distortion increases, the $t_{2g}$-$e_g$ hybridization ($t_3$) significantly increases while the hybridization between the neighboring $t_{2g}$ orbitals gradually decreases. Consequently, the energy $t_1^2j/{u'}^2 + t_3^2j/(u'+\Delta_{e_g})^2$ increases with increasing distortion while the energy $t_2^2/u$ decreases, leading to crossing of these two energies as shown in Fig~\ref{03fig06}(a). As a result, the spin-exchange along the $c$-axis changes from positive (AFM) to negative (FM) nearly continuously (see Fig~\ref{03fig06}(b)) We further note that since there exists a very slight difference between the orbital states in the AFM(A) and FM phases, a tiny jump of the spin exchanges should appear at the AFM(A)-FM phase boundary, resulting in the very weak first-order phase transition.
\begin{figure}[tdp]
\caption{Characteristic second-order perturbational energy gain due to
 the transfer processes along the $c$-axis for the antiparallel spin
 configuration $t_2^2/u$, and that for the parallel spin configuration
 $t_1^2j/{u'}^2 + t_3^2j/(u'+\Delta_{e_g})^2$ are plotted as functions of the Ti-O-Ti bond angle. Two energies are crossing at about 141$^{\circ}$. With decreasing Ti-O-Ti bond angle, the sign of the spin-exchange constant along the $c$-axis changes from positive (AFM) to negative (FM) continuously at the phase boundary. (From Mochizuki and Imada, Ref.~\cite{Mochizuki01a})}
\label{03fig06}
\end{figure}

Now, let us summarize the above discussion. In the compounds with strong GdFeO$_3$-type distortion, the ($yz$,$zx$,$yz$,$zx$)-type orbital ordering is strongly stabilized irrespective of the magnetic structure due to both the $d$-type Jahn-Teller distortion and the $t_{2g}$-$e_g$ hybridizations induced by the GdFeO$_3$-type distortion. Theoretically, if we assume this orbital ordering, the spin-exchange along the $c$-axis changes from AFM to FM with increasing GdFeO$_3$-type distortion due to the increased $t_{2g}$-$e_g$ hybridizations, resulting in the AFM(A)-FM phase transition. This indicates that for the emergence of the FM ordering in the perovskite titanates, the $e_g$ orbital degrees of freedom play a decisive role even though the titanates are the typical $t_{2g}$-electron system. In addition, a nearly two-dimensional spin coupling with $J_c \sim 0$ is realized at the AFM(A)-FM transition point. Hence, in the FM phase, a two-dimensional anisotropy increases as the GdFeO$_3$-type distortion decreases. The decrease of $T_{\rm C}$ as the system approaches the AFM-FM phase boundary in the phase diagram is attributed to this two-dimensional anisotropy. Through these discussions, it is shown that the GdFeO$_3$-type distortion, which has been considered as a bandwidth-control mechanism so far, controls the orbital-spin structures in the perovskite titanates by introducing the $t_{2g}$-$e_g$ hybridizations.

The importance of the $e_g$ orbitals can be easily understood as follows. The order of the spin exchange due to virtual transfer processes via the unoccupied $e_g$ orbitals is $t^2/(u'+\Delta_{e_g})$ with $t$ being a typical amplitude of the electron transfer between the Ti $3d$ orbitals. Here, the value of $\Delta_{e_g}$ is estimated as $\sim$1.7-2.0 eV from the optical measurements~\cite{Arima93,Higuchi99,Higuchi00,Higuchi03} and from the first-principles calculation~\cite{Bouarab96}. On the other hand, that via the singly-occupied $t_{2g}$ orbitals is $t^2/u=t^2/(u'+2j)$. Since $\Delta_{e_g}$ is comparable to $2j\sim 1.3$ eV, the $e_g$-orbital contribution $t^2/(u'+\Delta_{e_g})$ is not negligible, and actually plays a substantial role on the stability of the FM state in GdTiO$_3$ and YTiO$_3$.

Here, we note that the AFM ordering realized in the compounds with large $R$ ions such as La, Ce, Pr, Nd and Sm, and that in the La-rich solid solutions La$_{1-x}$Y$_x$TiO$_3$ is not A-type but G-type. In the above discussions, to focus on the FM compounds in the region of strong GdFeO$_3$-type distortion, the large $d$-type Jahn-Teller distortion is assumed, which was observed in GdTiO$_3$ and YTiO$_3$. Hence, the AFM(G) state in $R$TiO$_3$ ($R=$La, Pr, Nd and Sm) is beyond the scope of this framework, since in these compounds, the Jahn-Teller distortion is very small.

The present calculations, which assume the strong $d$-type Jahn-Teller distortion show that the AFM(A) state is expected when the strong $d$-type Jahn-Teller distortion is realized in the region of moderate GdFeO$_3$-type distortion. Here, ``moderate'' means ``not strong enough to make the system FM''. However, this condition is, in fact, hard to be realized because the $d$-type Jahn-Teller distortion becomes weak rapidly with decreasing GdFeO$_3$-type distortion. Indeed, SmTiO$_3$ ($\angle$TiOTi$=147^{\circ}$) does not show the $d$-type Jahn-Teller distortion while in GdTiO$_3$ ($\angle$TiOTi$=144^{\circ}$), a strong distortion is realized. There is no stoichiometric compound between SmTiO$_3$ (AFM(G), Ref.~\cite{Amow98}) and GdTiO$_3$ (FM) so that there is no possibility of the AFM(A) ordering in the stoichiometric systems. However, this AFM(A) state may possibly be realized in the solid-solution systems like La$_{1-x}$Y$_x$TiO$_3$ sandwiched by the AFM(G) and FM phases. The AFM(A) state may be found if we precisely control the magnitude of the GdFeO$_3$-type distortion by using the solid solutions or by applying pressures. However, it should be noted that even if the AFM(A) state is not realized in the actual systems, the validity of present discussion on the origin and nature of the FM state would never be affected.

\begin{figure}[tdp]
\caption{Spin-wave spectrum for YTiO$_3$ measured by inelastic
 neutron-scattering experiment. The closed symbols refer to the magnetic excitations, while the open symbols are identified as phonons. (From Ulrich $et$ $al.$, Ref.~\cite{Ulrich02})}
\label{03fig07}
\end{figure}
Finally, we mention a recent puzzling finding from the inelastic neutron-scattering experiment by Ulrich $et$ $al.$~\cite{Ulrich02}. They reported that the spin-wave spectrum of YTiO$_3$ is well described by the isotropic FM Heisenberg model (see Fig.~\ref{03fig07}). On the other hand, if we take the ($yz$,$zx$,$yz$,$zx$)-type orbital ordering model, the FM coupling should have a two-dimensional anisotropy. This discrepancy has not been clearly resolved yet. To explain the isotropic FM exchanges, Khaliullin and Okamoto recently proposed a certain type of orbital ordering model~\cite{Khaliullin02,Khaliullin03}. This proposal is based on the analysis of a model with the triply degenerate $t_{2g}$ orbitals on the cubic lattice, and the effects of the actual lattice distortions (the GdFeO$_3$-type distortion and the Jahn-Teller distortion) are not taken into account. In fact, the predicted orbital structure contradicts the experimentally observed one. This contradiction may be attributed to the neglect of the lattice distortions, and may imply that it is necessary to take into account the lattice distortions when we consider the orbital-spin structures in the FM titanates as we have discussed in this section. This problem should be clarified by further studies in the future. Considering the fact that the energy scale of the phonon spectrum is comparable to that of the magnon spectrum, it may be necessary to consider the orbital excitations coupled with the lattice fluctuations.

\section{AFM(G) ordering and orbital structure in LaTiO$_3$}

LaTiO$_3$ with the smallest GdFeO$_3$-type distortion exhibits the AFM(G) ordering with $T_{\rm N}\sim$140-150 K. The magnetic ordered moment is 0.57 $\mu_{\rm B}$~\cite{Cwik03}, which is strongly reduced from spin-1/2 moment of 1 $\mu_{\rm B}$. Although the orbital-spin structure in LaTiO$_3$ has been studied intensively, even the origin of this AFM(G) ordering had not been elucidated. However, a recent theoretical study showed that the emergence of the AFM(G) ordering and puzzling experimental findings are consistently explained when we consider the crystal field from the La ions induced by the GdFeO$_3$-type distortion~\cite{Mochizuki03,Mochizuki04}. In this section, we would like to discuss the orbital-spin structure and several experimental results of LaTiO$_3$ on the basis of this theoretical proposal.

Now, we first briefly summarize the history of studies and controversies on the orbital-spin structure in LaTiO$_3$. Previous diffraction studies showed that a Jahn-Teller type distortion of the TiO$_6$ octahedron in LaTiO$_3$ is undetectably small~\cite{MacLean79,Eitel86}. If this is true, the crystal field from the O ions surrounding the Ti$^{3+}$ ion has a cubic symmetry so that the $t_{2g}$-orbital degeneracy is expected to survive.

Under this circumstance, the emergence of the AFM(G) ordering had been surprising since in the orbitally degenerate systems, it is theoretically expected that a FM state with antiferro-orbital ordering is favored both by the electron transfers and by the Hund's-rule coupling~\cite{Roth66,Kugel72,Kugel73,Inagaki73,Cyrot73,Cyrot75}. Indeed, a recent theoretical study based on a Kugel-Khomskii model showed that in the cubic lattice with threefold degenerate $t_{2g}$ orbitals, a FM state is almost always stable, and the AFM(G) state appears only in the unphysical regions of spin-exchange parameters~\cite{Ishihara02}.

On the basis of the Hartree-Fock analysis of the multiband $d$-$p$ model, Mizokawa and Fujimori proposed that the spin-orbit (LS) interaction lifts the $t_{2g}$ degeneracy, and the AFM(G) state in LaTiO$_3$ is accompanied by the LS ground state, in which two states $\frac{1}{\sqrt{2}}(y'z'+iz'x')\uparrow$ and $\frac{1}{\sqrt{2}}(y'z'-iz'x')\downarrow$ with $z'$-axis and spins pointing in the (1,1,1) direction are alternating between the neighboring Ti sites~\cite{Mizokawa96a,Mizokawa96b}. In the LS ground state, unquenched orbital moment antiparallel to the spin moment is necessarily induced. At first sight, this seemed to be consistent with the experimentally observed reduced magnetic moment in this compound.

\begin{figure}[tdp]
\caption{Spin-wave dispersion in the (1,1,1) direction of the reciprocal space for LaTiO$_3$ measured by neutron-scattering experiment. The line is the fitting curve by using the Heisenberg model with isotropic $J$ of 15.5 meV on the cubic lattice. (From Keimer $et$ $al.$, Ref.~\cite{Keimer00})}
\label{04fig01}
\end{figure}
However, in contrast with this naive prediction, a recent neutron scattering study showed a spin-wave spectrum well described by the isotropic spin-1/2 Heisenberg model ($J\sim$15.5 meV) with a considerably small spin gap of $\sim$3 meV (see Fig.~\ref{04fig01})~\cite{Keimer00}. This indicates that the orbital moment is almost quenched, and the LS interaction is irrelevant since if the orbital moment exists, an induced Ising-type anisotropy in the spin sector has to generate a large spin gap. Actually, an exact diagonalization study showed that the Kugel-Khomskii model with the LS interaction does not describe the observed spin-wave spectrum~\cite{Miyahara02}. In addition, a model Hartree-Fock study done by Mochizuki~\cite{Mochizuki02} also showed that the LS ground state cannot reproduce the AFM(G) ordering by examining a solution which was overlooked in the previous Mizokawa and Fujimori's studies of Refs.~\cite{Mizokawa96a,Mizokawa96b}. In Ref.~\cite{Mochizuki02}, it was shown that even without the static Jahn-Teller distortion, a FM state out of which two states $\frac{1}{\sqrt{2}}(yz+izx)\uparrow$ and $xy\uparrow$ are alternating is stabilized both by the LS interaction and by the spin-orbital superexchange interaction.

To explain the neutron scattering result, a possible orbital liquid state was proposed on the basis of small orbital-exchange interaction in the AFM(G) spin structure~\cite{Khaliullin00,Khaliullin01,Kikoin03}. In these theories, the AFM(G) ordering is assumed a priori. However, in the titanates, the spins and orbitals strongly couple each other, and both degrees of freedom cannot be determined independently. Therefore, the origin of the AFM(G) state in LaTiO$_3$ is to be clarified in a self-consistent manner. More importantly, a FM state with antiferro-orbital ordering was theoretically expected to be more stable in this system. Thus, the assumption of the AFM(G) order on this basis is hard to be justified since relative stability of the AFM(G) to FM state has never been examined. In addition, it has been recently proven that the Kugel-Khomskii model with triply degenerate $t_{2g}$ orbitals and zero Hund's-rule exchange on the undistorted cubic lattice, on which the orbital liquid theory is based does not exhibit the long-range magnetic ordering at any finite temperature~\cite{Harris03,Harris04}. This also questions a validity of the orbital liquid theory since this theory assumes the AFM(G) ordering in the limit of weak Hund's-rule coupling. Indeed, a recent heat capacity measurement showed that the most of the low-temperature heat capacity arises from magnon contributions, which contradicts the prediction from the orbital liquid theory about the orbital contributions~\cite{Fritsch02}.

A Raman scattering measurement performed for half metallic La$_{1-y}$TiO$_3$ with slight La-ion vacancies observed a Fano-type anomaly in the phonon spectrum~\cite{Reedyk97}. The origin of this anomaly was identified as the coupling of phonons and orbital excitations in the orbital liquid theory~\cite{Khaliullin00,Khaliullin01}, and this observation was considered to support this theory. However, another Raman scattering measurement performed for stoichiometric LaTiO$_3$ did not observe the anomaly~\cite{Iliev03}.

\begin{figure}[tdp]
\caption{With the $D_{3d}$ distortion, the TiO$_6$ octahedron is contracted along a trigonal direction. There exist two kinds of O-O bonds, shorter O-O bond and longer O-O bond as presented by solid lines and dashed lines, respectively. As a result of this distortion, the threefold degenerate $t_{2g}$ levels split into a nondegenerate lower $a_{1g}$ level and twofold degenerate $e_g$ levels. With the $D_{3d}$ crystal field with [1,1,1] trigonal axis, the representation of the $a_{1g}$ orbital is $\frac{1}{\sqrt{3}}(xy+yz+zx)$.}
\label{04fig02}
\end{figure}
A possible trigonal ($D_{3d}$) distortion of the TiO$_6$ octahedra was proposed to explain the emergence of the AFM(G) structure~\cite{Mochizuki01b}. With this distortion, the TiO$_6$ octahedron is contracted along the threefold direction, and the threefold degenerate $t_{2g}$ level splits into a nondegenerate lower $a_{1g}$ level and twofold-degenerate higher $e_g$ level (see Fig.~\ref{04fig02}). Occupations of the $a_{1g}$ orbitals well explain the emergence of the AFM(G) ordering and the isotropic spin-wave spectrum. However, this distortion had not been observed clearly upon theoretical proposal.

NMR measurements were also applied to this compound. Previously, the NMR spectrum was analyzed based on the orbital liquid picture~\cite{Itoh99}. However, it was recently claimed that there exists a discrepancy in this analysis. More specifically, the NMR spectra show the large quadrupole moment of a $3d$ electron in LaTiO$_3$, which is unfavorable for the orbital liquid model. Instead, it was recently clarified that the spectrum is well described by an orbital ordering model proposed in the above $D_{3d}$-distortion scenario instead of the orbital liquid model~\cite{Kiyama02}. Moreover, a recent resonant x-ray scattering result also indicates the orbital orderings in the series of the AFM(G) compounds of $R$TiO$_3$ ($R=$La, Pr, Nd and Sm), which have the same symmetry with the orbital orderings observed in the FM compounds, GdTiO$_3$ and YTiO$_3$~\cite{Kubota00}. This also contradicts the orbital liquid picture. 

Several first-principles methods were also applied to LaTiO$_3$. However, LDA and LSDA failed to reproduce not only AFM(G) structure but also insulating behavior since the strong electron correlation is insufficiently treated~\cite{Pari95,Mahadevan96,Solovyev96,Hamada97}. In addition, although the insulating gap was reproduced, the AFM(G) structure was not obtained in the LDA+$U$ calculation~\cite{Solovyev96}. These indicate that in this compound, the effect of strong electron correlations is crucially important, which is of general interest of the condensed matter physics. 

In such ways as discussed above, the problem of the orbital-spin structure in LaTiO$_3$ had been a subject of hot debates, but a consistent theory had not emerged. In a number of previous studies, the degeneracy of the $t_{2g}$ orbitals had been assumed since the TiO$_6$ octahedra retain the cubic symmetry. However, this assumption is, in fact, a source of controversies. A recent theoretical study showed that the $t_{2g}$ degeneracy is actually lifted by the crystal field from the displaced La ions in the GdFeO$_3$-type structure~\cite{Mochizuki03,Mochizuki04}. The orbital-spin structure and available experimental results can be well explained if we consider the effects of this La crystal field.

\begin{figure}[tdp]
\caption{(a) Displacements of the La cations around a TiO$_6$ octahedron (site 1) in the GdFeO$_3$ structure are presented by arrows. In LaO-plane 1, they shift in the negative direction along the $b$ axis while in LaO-plane 2, they shift in the positive direction. The distances between the Ti ion and two La ions (gray circles) located in the $\pm(1,1,1)$-directions are decreased. As a result, the La ions generate a crystal field similar to the $D_{3d}$ crystal field with [1,1,1] trigonal axis (see text). Note that rotation of the octahedron due to the GdFeO$_3$-type distortion is not presented. (b) Stacking of the TiO$_6$ octahedra as well as shifts of the La ions are presented. Note that Ti and O ions are not explicitly presented in this figure. (From Mochizuki and Imada, Ref.~\cite{Mochizuki03})}
\label{04fig03}
\end{figure}
Now, according to Refs.~\cite{Mochizuki03,Mochizuki04}, let us discuss the effects of the crystal field from the La ions, which are displaced by the GdFeO$_3$-type distortion in LaTiO$_3$. With the GdFeO$_3$-type distortion, the La ions shift mainly along the $(1,1,0)$-axis (namely, the $b$-axis), and slightly along the $(1,-1,0)$-axis (the $a$-axis)~\cite{MacLean79,Eitel86,Mizokawa99}. There are two kinds of LaO-planes (plane 1 and plane 2) stacking alternatingly along the $c$-axis ($(0,0,1)$-axis) as shown in Fig.~\ref{04fig03}. In plane 1, the La ions shift in the negative direction along the $b$-axis while they shift in the positive direction in plane 2. Then, distances between Ti and La ions are no longer the same. For sites 1 and 2, the distances between the Ti ion and the La ions located in the $\pm(1,1,1)$-directions decrease while those along the $\pm(1,1,-1)$-directions increase (see Fig.~\ref{04fig03} (a)). On the other hand, the distances along $\pm(1,1,-1)$-directions decrease while those along  $\pm(1,1,1)$-directions increase for sites 3 and 4. (In Fig.~\ref{04fig03} (b), the stacking of each site is presented.) The changes of the other Ti-La distances are rather small. As a result, the crystal field from the La ions is distorted from a cubic symmetry.

On the other hand, without any distortion of the TiO$_6$ octahedron, the crystal field from the ligand oxygens has a cubic symmetry. Under this circumstance, when we introduce the crystal field from the La ions, the threefold degeneracy of the cubic-$t_{2g}$ level splits. We expect that the La ions generate an attractive crystal-field potential and the Ti $3d$ orbitals directed along the shorter Ti-La bonds are lowered in energy because the nominal valence of the rare-earth ion is 3+. Therefore, we expect that the lowest orbitals are approximately $\frac{1}{\sqrt{3}}(xy+yz+zx)$, $\frac{1}{\sqrt{3}}(xy+yz+zx)$, $\frac{1}{\sqrt{3}}(xy-yz-zx)$ and $\frac{1}{\sqrt{3}}(xy-yz-zx)$ at sites 1,2,3 and 4, respectively.

In order to confirm this, calculations based on the point charge model was performed by assuming the +3 valence on each La ion. In this calculation, the Coulomb interaction between an electron on a Ti $3d$ orbital and surrounding La$^{3+}$ ions is given by using a dielectric constant $\epsilon_{\rm TiLa}$ as,
\begin{equation}
 v(\vct {$r$})=-\sum_{i}
 \frac{Z_R e^2}{\epsilon_{\rm TiLa}|{\vct {$R$}}_i - {\vct {$r$}}|},
\end{equation}
where $\vct {$R$}_i$ expresses the coordinate of the $i$-th La ion, which is deduced from the x-ray diffraction data~\cite{MacLean79}, and $Z_R$(=+3) is the nominal valence of the La$^{3+}$ ion.

The calculation showed that due to the attractive Coulomb potential from the La$^{3+}$ ions, the $t_{2g}$ level splits into three isolated levels. The wavefunctions of the lowest orbitals at each site are specified by the linear combinations of the $xy$, $yz$ and $zx$ orbitals as follows,
\begin{eqnarray}
&{\rm site}&\quad 1; \quad a|xy\rangle+c|yz\rangle+b|zx\rangle, \nonumber \\
&{\rm site}&\quad 2; \quad a|xy\rangle+b|yz\rangle+c|zx\rangle, \nonumber \\
&{\rm site}&\quad 3; \quad a|xy\rangle-c|yz\rangle-b|zx\rangle, \nonumber \\
&{\rm site}&\quad 4; \quad a|xy\rangle-b|yz\rangle-c|zx\rangle.
\label{eqn:orbstrct1}
\end{eqnarray}
where $a^2$+$b^2$+$c^2$=1. The coefficients take $a$=0.60, $b$=0.39 and $c$=0.69. Here, the $xy$, $yz$ and $zx$ orbitals are defined in terms of the $x$-, $y$-, and $z$-axes attached to each TiO$_6$ octahedra. The NMR spectrum proved to be well described by this orbital ordering model~\cite{Kiyama03}. In addition, dynamical mean-field calculations combined with first principles local density approximation (LDA+DMFT) showed similar orbital ordering with nondegenerate orbital ground state for LaTiO$_3$~\cite{Pavarini03,Craco03}.

\begin{figure}[tdp]
\caption{Orbital structure in the $D_{3d}$ crystal fields with [1,1,1], [1,1,1], [1,1,$-$1] and [1,1,$-$1] trigonal axes at sites 1, 2, 3 and 4, respectively. The orbital wavefunctions at each site are $\frac{1}{\sqrt{3}}(xy+yz+zx)$, $\frac{1}{\sqrt{3}}(xy+yz+zx)$, $\frac{1}{\sqrt{3}}(xy-yz-zx)$ and $\frac{1}{\sqrt{3}}(xy-yz-zx)$, respectively. Note that tiltings due to the GdFeO$_3$-type distortion are not presented explicitly in this figure, and the orbital basis $xy$, $yz$ and $zx$ are defined in terms of the $x$-, $y$- and $z$-axes attached to the tilted octahedron in reality.}
\label{04fig04}
\end{figure}
In this orbital ordering, the orbital wavefunctions at sites 1 and 2 are actually similar to $\frac{1}{\sqrt{3}}(xy+yz+zx)$, and at sites 3 and 4 they are similar to $\frac{1}{\sqrt{3}}(xy-yz-zx)$. In fact, $\frac{1}{\sqrt{3}}(xy+yz+zx)$ is the lowest orbital for the $D_{3d}$ crystal field with [1,1,1] trigonal axis, while $\frac{1}{\sqrt{3}}(xy-yz-zx)$ is the lowest orbital for that with [1,1,$-$1] trigonal axis. The orbital structure realized in the La crystal field is similar to that with the lowest orbitals in the $D_{3d}$ crystal fields with [1,1,1], [1,1,1], [1,1,$-$1] and [1,1,$-$1] trigonal axes at sites 1, 2, 3 and 4, respectively (see Fig.~\ref{04fig04}). Therefore, this orbital ordering is expected to induce the $D_{3d}$ distortions of the unit TiO$_6$ octahedra with these trigonal axes. In fact, the possibility of the $D_{3d}$ distortions in LaTiO$_3$ was previously proposed in Ref.~\cite{Mochizuki01b}. After this proposal, a number of diffraction studies were performed~\cite{Arao02,Hemberger03,Cwik03}, and such distortions were actually detected by an x-ray diffraction measurement~\cite{Cwik03}. 

One might argue that this TiO$_6$ distortion could occur by the usual Jahn-Teller mechanism. However, this is not true because the orbital degeneracy is already lifted by the La crystal field and the Jahn-Teller mechanism is not effective if the GdFeO$_3$-type distortion (tilting of the TiO$_6$ octahedra) is present. The observed TiO$_6$ distortion is simply induced by the orbital ordering due to the La crystal field. In this sense, the $t_{2g}$ degeneracy is lifted mainly by the La crystal field, and the TiO$_6$ distortion occurs subsequently.

Here, we note that the crystal field from the La ions has two origins. One is the Coulomb potential from the charged La ions as we have discussed above, and the other is the hybridizations between the Ti $3d$ orbitals and unoccupied orbitals on the La ions. The latter effect was also examined by using the second-order perturbational expansion in terms of the electron transfers between the Ti $3d$ and La $5d$ orbitals. The Slater-Koster parameters for the Ti $3d$-La $5d$ hybridizations and the energy difference between these two orbitals are deduced from the photoemission spectra~\cite{Fujimori92a} and the LDA band calculation~\cite{Hamada02}, respectively. It was shown that the hybridizations between the Ti $3d$ and La $5d$ orbitals also stabilize the orbital ordering in Eq.~(\ref{eqn:orbstrct1}), and works cooperatively with the Coulomb attractive potential.

We next discuss the stability of the magnetic state in the La crystal field (Hereafter, the Hamiltonian for the La crystal field is referred to as $H_{\rm La}$). An energy calculation by using the effective spin-pseudospin Hamiltonian introduced in Sec. 3 was performed. Substituting the crystal field term $H_{\rm cry.}$ with $H_{\rm La}$, the energies for several magnetic structures were calculated by applying the mean-field approximation. This calculation showed that the AFM(G) ordering is the most stable solution.

 The spin-exchange constants were also calculated. Since the level splitting due to the La crystal field is much larger than $k_{\rm B}T_{\rm N}$, the orbital occupation is restricted to the lowest orbitals irrespective of the spin structure. Then, we can estimate the magnitude of the spin-exchange interaction in the subspace of singly-occupied lowest orbitals. The spin-exchange constant $J$ for a Ti-Ti bond is represented as,
$J=(E_{\uparrow\uparrow}-E_{\uparrow\downarrow})/2S^2$
with $E_{\uparrow\uparrow}$ and $E_{\uparrow\downarrow}$ being mean-field energy gains for the Ti-Ti bond of $\uparrow\uparrow$- and $\uparrow\downarrow$-pairs, respectively. For LaTiO$_3$, the values of $J$ along the $x$-, $y$- and $z$-axes take as $J_x$=18.5 meV, $J_y$=18.5 meV and $J_z$=19.7 meV, respectively, which well reproduce the spin-wave spectrum with isotropic exchange constant of $\sim$15.5 meV. 

In addition, the $t_{2g}$ splitting is sufficiently larger than the coupling constant of the LS interaction in Ti$^{3+}$ (${\zeta}_d$=0.018 eV)~\cite{Sugano70} so that the La crystal field dominates over the LS interaction, resulting in the quenched orbital moment. This is also consistent with the observed small magnon gap in the spin-wave spectrum.

As discussed above, by considering the $t_{2g}$-level splitting due to the La crystal field, the emergence of the AFM(G) ordering and the spin-wave spectrum characterized by (1) isotropic AFM coupling, (2) $J\sim15.5$ meV, and (3)very small spin gap can be well explained.

Here we mention that for more detailed description of the magnetic structure in LaTiO$_3$, it is necessary to consider the Dzyaloshinskii-Moriya (DM) interaction. For example, the AFM(G) ordering in this compound is accompanied by a weak FM moment due to spin canting~\cite{Goral83}, and this weak ferromagnetism is explained by the DM interaction~\cite{Schmitz04}. In addition, the small magnon gap in the spin-wave spectrum is also attributed by this interaction~\cite{Keimer00}. However, the origin and overall features of the magnetic structure in LaTiO$_3$ are well explained even without this interaction.

Now, let us discuss a problem of the reduced ordered moment in LaTiO$_3$. The ordered moment was measured as 0.57 $\mu_{\rm B}$ in the neutron-diffraction study~\cite{Cwik03}, or as 0.45 $\mu_{\rm B}$ in the earlier powder neutron-diffraction study~\cite{Goral83}. These value had been considered to be much smaller than 0.85 $\mu_{\rm B}$ of the spin-wave theory on the 3D Heisenberg model. This reduction was attributed to the antiparallel contribution of the orbital moment due to the LS interaction. However, as discussed above, the recent neutron-scattering experiment showed that the orbital moment is fully quenched. Then, the origin of the moment reduction has been recognized as a puzzle. However, this is not so puzzling actually when we consider the strong charge and orbital fluctuations in LaTiO$_3$.

In fact, LaTiO$_3$ has a small insulating gap of 0.2 eV, and this compound is located in the vicinity of the metal-insulator transition point~\cite{Arima93,Okimoto95}. Thus, in LaTiO$_3$, a large amount of itinerant fluctuations of charges and orbitals should remain. Although these fluctuations are completely neglected in the Heisenberg model in the insulating limit (in the limit of vanishing these fluctuations), we expect that in the actual compound, the fluctuations strongly reduce the magnetic moment. For example, a quantum Monte-Carlo simulation performed by White $et$ $al.$ showed that the charge fluctuations diminish the ordered moment of $\sim$0.6 $\mu_{\rm B}$ for the 2D Heisenberg model to $\sim$0.35 $\mu_{\rm B}$ for the 2D Hubbard model with $U=4t$~\cite{White89}. In addition to the charge fluctuations, the imperfect nesting of the Fermi surface and the orbital fluctuations also work as reducing the magnetic moment~\cite{Kakehashi86,Kashima01,Oles83}, and the observed reduced moment can be understood by considering these effects as discussed in Ref.~\cite{Mochizuki04} in detail.

We note that the insulating behavior and the magnetic properties such as magnetization and $T_{\rm N}$ for LaTiO$_3$ are highly sensitive to the slight off-stoichiometry induced by the excess oxygens or by the La-ion vacancies~\cite{Lichtenberg91,Crandles94,Meijer99} Indeed, the early measurement using a poorly stoichiometric sample observed the magnetic moment of 0.45 $\mu_{\rm B}$~\cite{Goral83}, which is smaller than the recently reported value of 0.57 $\mu_{\rm B}$~\cite{Cwik03}. This difference may be attributed to the induced small amount of carriers, which cause the itinerant fluctuations.

In this section, we have discussed controversies on the orbital-spin structure in LaTiO$_3$. The La crystal field generated by the GdFeO$_3$-type distortion lifts the Ti $t_{2g}$-orbital degeneracy. This leads to the orbital ordering in the ground state instead of the orbital liquid, which provides a consistent description for the emergence of the AFM(G) ordering and the puzzling experimental results. As we have discussed, the following experimental findings for LaTiO$_3$ are explained: (1) the spin-wave spectrum with isotropic $J$ of $\sim15.5$ meV and small magnon gap measured by the neutron-scattering experiment, (2) the orbital structure detected by the NMR experiment, and (3) the trigonal distortions of the TiO$_6$ octahedra observed in the x-ray diffraction measurement. In addition, a quite recent spin-resolved photoelectron spectroscopic experiment with circular polarized light showed that in LaTiO$_3$, the crystal field splitting of the $t_{2g}$ subshell is about 0.12-0.30 eV and the orbital moment is strongly reduced~\cite{Haverkort04}. This is also consistent with the present theory.

In the previous section, it has been shown that the GdFeO$_3$-type distortion has a universal relevance on the orbital-spin structures in the titanates through inducing the $t_{2g}$-$e_g$ hybridizations. Now, we mention that the GdFeO$_3$-type distortion controls the orbital-spin structures also through generating the $R$ crystal field. It should be noted that this mechanism is not a usual Jahn-Teller one. In the Jahn-Teller systems, 3$d$ electrons on the transition-metal ions and the lattice degree of freedom strongly couple each other, and the lattice spontaneously distorts to lift the 3$d$ degeneracy and lower the electronic energy. On the other hand, the GdFeO$_3$-type distortion is caused by interactions between the $R$ ions and O ions. More concretely, the octahedral tilting is induced by energy gain in $R$-O covalency bonding in the perovskite compounds with $R$-site mismatching, and subsequent $R$-site shifts occur by $R$-O ionic interactions~\cite{Woodward97}. This distortion exists irrespective of the 3$d$ electronic state, and has nothing to do with any spontaneous lift of the electronic degeneracy in contrast with the Jahn-Teller mechanism. The lifting of the degeneracy of the Ti 3$d$ orbitals occurs just as a consequence of the crystal field from the La ions. In this sense, the mechanism here proposed is different from the Jahn-Teller, which is inherent and universal with the GdFeO$_3$-type distortion. In fact, this mechanism also works in other AFM(G) titanates with $R=$Ce, Pr, Nd and Sm as we will discuss in the next section.

\section{Orbital-spin structures in $R$TiO$_3$ ($R$=La, Ce, Pr, Nd and Sm)}

In the previous section, we have discussed that the effect of the La crystal field consistently explains the puzzling experimental results as well as the emergence of the AFM(G) ordering in LaTiO$_3$. In fact, it was shown that the crystal fields from the $R$ ions also play important roles in $R$TiO$_3$ with other $R$ ions~\cite{Mochizuki03,Mochizuki04}. In this section, we discuss the orbital-spin structures in $R$TiO$_3$ with $R$ being La, Ce, Pr, Nd and Sm by considering the effects of the $R$ crystal field.

\begin{figure}[tdp]
\caption{Orbital wavefunctions at each Ti site are presented as functions of the Ti-O-Ti bond angle. The dotted line indicates 1/$\sqrt{3}$. Open symbols at 140$^{\circ}$ represent the orbital wavefunction of YTiO$_3$ obtained by polarized neutron-scattering experiments ($a$=$\sqrt{0.4}$ and $c$=$\sqrt{0.6}$, Ref.~\cite{Ichikawa00,Akimitsu01}). When the bond angle decreases, the orbitals continuously approach the character of YTiO$_3$.}
\label{05fig01}
\end{figure}
First, let us discuss the orbital state in the $R$ crystal field of these compounds. The crystal field Hamiltonians based on the point charge model for these compounds were derived again by using the experimentally measured coordination parameters, and the wavefunctions of the lowest orbitals were calculated. The lowest orbitals for $H_{R1}$ of these compounds are expressed also by Eq.~(\ref{eqn:orbstrct1}). In Fig.~\ref{05fig01}, we show the coefficients $a$, $b$ and $c$ as functions of the Ti-O-Ti bond angle. In LaTiO$_3$ with the smallest GdFeO$_3$-type distortion, the orbital wavefunctions are similar to those of the lowest orbitals in the $D_{3d}$ crystal fields, namely $\frac{1}{\sqrt{3}}(xy+yz+zx)$ or $\frac{1}{\sqrt{3}}(xy-yz-zx)$. The weights of the $xy$, $yz$ and $zx$ orbitals are nearly the same in LaTiO$_3$. 

As the GdFeO$_3$-type distortion increases, the orbital wavefunctions become to differ from those of LaTiO$_3$ since the locations of the $R$ ions gradually deviate from the trigonal directions due to the increased tiltings. With increasing GdFeO$_3$-type distortion, the component of the $zx$ orbital monotonically decreases at sites 1 and 3, while the $yz$-component decreases at sites 2 and 4. The reductions of these orbital occupations indicate that with increasing GdFeO$_3$-type distortion, the crystal field from the $R$ cations tends to progressively stabilize the orbital structure in which sites 1, 2, 3  and 4 are approximately occupied by the orbitals as follows:
\begin{eqnarray}
&{\rm site}&\quad 1; \quad
\frac{1}{\sqrt{2}}(xy+yz), \nonumber \\
&{\rm site}&\quad 2; \quad  
\frac{1}{\sqrt{2}}(xy+zx), \nonumber \\
&{\rm site}&\quad 3; \quad 
\frac{1}{\sqrt{2}}(xy-yz), \nonumber \\
&{\rm site}&\quad 4; \quad 
\frac{1}{\sqrt{2}}(xy-zx).
\label{eqn:orbstrct2}
\end{eqnarray}

In fact, this orbital structure is the same as the orbital ordering which has been observed in YTiO$_3$ with large $d$-type Jahn-Teller distortion by several experiments~\cite{Itoh97,Itoh99,Itoh01,Ichikawa00,Akimitsu01,Nakao02}. More concretely, $a$ and $c$ seem to be extrapolated to values obtained by a polarized neutron study~\cite{Ichikawa00,Akimitsu01} ($a$=$\sqrt{0.4}$ and $c$=$\sqrt{0.6}$). Further, a recent resonant x-ray scattering study showed that the orbital state continuously changes as $R$ goes from La to Y, and both the orbital state in SmTiO$_3$ (AFM(G)) and that in GdTiO$_3$ (FM) have the same twofold symmetry, indicating that the symmetry of the orbital structure does not change through the magnetic phase transition~\cite{Kubota00}. This experimental finding is consistent with the present theoretical result.

Now, we mention a relation between the TiO$_6$ distortion and the $R$ ions. In the previous section, we have mentioned that in LaTiO$_3$, the orbital ordering stabilized by the La crystal field favors the stacking of the trigonally distorted TiO$_6$ octahedra with [1,1,1] and [1,1,$-$1] axes alternatingly along the $c$-axis, which was actually detected experimentally. On the other hand, GdTiO$_3$ and YTiO$_3$ exhibit the $d$-type Jahn-Teller distortion in which the elongated axes of the octahedra are parallel in the $c$-plane as shown in Fig.~\ref{03fig01}. As a consequence of this distortion, the $xy$ and $yz$ orbitals are lowered in energy at sites 1 and 3, while at sites 2 and 4, the $xy$ and $zx$ orbitals are lowered. In fact, the orbital wavefunctions in Eq.~(\ref{eqn:orbstrct2}) are represented by linear combinations of the twofold lower orbitals in the $d$-type Jahn-Teller distortion. This suggests that the $d$-type Jahn-Teller distortion in GdTiO$_3$ and YTiO$_3$ is favored by the orbital ordering due the $R$ crystal field. On the other hand, a recent multiband $d$-$p$ model study showed that with strong GdFeO$_3$-type distortion, the covalency between the O and $R$ ions stabilizes the $d$-type Jahn-Teller distortion~\cite{Mizokawa99}. These two effects, namely, the orbital ordering due to the $R$ crystal field and the $R$-O covalency may cooperatively work to stabilize the $d$-type Jahn-Teller distortion. It is interesting to point out that in the titanates, the $R$ ions in the GdFeO$_3$-type structure may control the distortion of the unit TiO$_6$ octahedron.

We also mention that the orbital structure in SmTiO$_3$ may slightly deviate from the orbitals presented here because of an additional effect of the O crystal field due to the relatively strong TiO$_6$ distortion. More concretely, we expect that in SmTiO$_3$, the occupation of the $xy$ orbital is somewhat increased at all Ti sites as will be discussed later. 

\begin{figure}[tdp]
\caption{Spin-exchange constants along the $x$-, $y$- and $z$-axes are plotted as functions of the Ti-O-Ti bond angle. (From Mochizuki and Imada, Ref.~\cite{Mochizuki04})}
\label{05fig02}
\end{figure}
We next discuss the magnetic states in $R$TiO$_3$ with $R$ being La, Ce, Pr, Nd and Sm. The AFM(G) spin structure is also reproduced in these compounds. In Fig.~\ref{05fig02}, the calculated spin-exchange constants $J_x$, $J_y$ and $J_z$ are plotted as functions of the Ti-O-Ti bond angle. This figure shows that as the GdFeO$_3$-type distortion increases, the spin exchange gradually decreases, which is consistent with the gradual decrease of $T_{\rm N}$ as $R$ goes from La to Sm.

The Neel temperature $T_{\rm N}$ for SmTiO$_3$ is about 50 K. This is 40$\%$ smaller than $T_{\rm N}$ for LaTiO$_3$. However, the calculated spin-exchange constant $J_x$ for SmTiO$_3$ is as large as 70$\%$ of that for LaTiO$_3$. This is puzzling when we consider the fact that $T_{\rm N}$ is proportional to $J$~\cite{Sandvik99} in the Heisenberg models.

This discrepancy can be solved by considering the additional effect of the O crystal field. In the above analysis, the $R$ crystal field due to the GdFeO$_3$-type distortion is examined by assuming only the cubic O crystal field. However, the TiO$_6$ distortion in SmTiO$_3$ is relatively strong so that we expect a substantial contribution from the O crystal field to the orbital-spin structure.

When the TiO$_6$ octahedron is strongly distorted from the cubic symmetry, the ligand oxygens are responsible for lifting of the orbital degeneracy through generating the crystal field. On the other hand, when the atomic TiO$_6$ distortion is small, it is expected that the $R$ crystal field plays a substantial role on lifting of the $t_{2g}$ and $e_g$ degeneracy, and dominantly determines the orbital-spin structures. Here, if the situation is inbetween, a competition between the $R$ crystal field and the O crystal field occurs.

The magnitude of the TiO$_6$ distortion is expressed by the standard deviation from the mean for the three inequivalent Ti-O bond lengths ($\sigma$).
\begin{figure}[tdp]
\caption{Standard deviation from the mean for three inequivalent Ti-O bond lengths is plotted as a function of the Ti-O-Ti bond angle.The data of the Ti-O bond lengths are deduced from Ref.~\cite{MacLean79}}
\label{05fig03}
\end{figure}
In Fig.~\ref{05fig03}, $\sigma$ obtained by the x-ray diffraction study~\cite{MacLean79} is plotted as a function of the Ti-O-Ti bond angle. The value for PrTiO$_3$ is deduced by the interpolation of LaTiO$_3$ and NdTiO$_3$. This figure shows that the values for $R$=La, Pr and Nd are rather small, indicating relatively small TiO$_6$ distortions. Thus, the $R$ crystal field actually determines the orbital-spin structures in these compounds. On the other hand, the values for $R$=Gd and Y are rather large because of the strong $d$-type Jahn -Teller distortions. As for SmTiO$_3$, the TiO$_6$ octahedra are moderately distorted, which leads to strong competition between the crystal field from the Sm ions and that from the O ions. In fact, the orbital-spin structure in SmTiO$_3$ has a two dimensional character due to this competition as discussed in the following.

\begin{figure}[tdp]
\caption{(a) Energy-level structures of the crystal field Hamiltonians $H_{L1}$ (upper panel) and $H_{L2}$ (lower panel) as functions of the Ti-O-Ti bond angle (see text). (b) TiO bonds in SmTiO$_3$ are elongated in the $xy$-plane, and are contracted along the $z$-axis. The data of the Ti-O bond lengths are taken from Ref.~\cite{MacLean79}. (From Mochizuki and Imada, Ref.~\cite{Mochizuki04})}
\label{05fig04}
\end{figure}
The $t_{2g}$ level splittings due to the O crystal fields in $R$TiO$_3$ were estimated. The O crystal field Hamiltonians of the electrostatic origin ($H_{L1}$) were derived by considering the Coulomb repulsive interaction between an electron on the Ti $t_{2g}$ orbitals and the ligand O$^{2-}$ ions within the point charge approximation. In addition, the O crystal field Hamiltonians for the hybridizations between the Ti $t_{2g}$ and O $2p$ orbitals ($H_{L2}$) were also derived. Here, in order to study the effects of the TiO-bond length variation, the O-Ti-O bond angle is assumed to be 90$^{\circ}$. In Fig.~\ref{05fig04} (a), the relative energies of each level for $H_{L1}$ (upper panel) and those for $H_{L2}$ (lower panel) are plotted. Here, $\epsilon_{\rm TiO}$ is a dielectric constant. In the compounds with small GdFeO$_3$-type distortion ($R=$La, Pr and Nd), the differences of the energies for each level are rather small whereas in largely distorted GdTiO$_3$ and YTiO$_3$, the $xy$ and $yz$ orbitals are strongly lowered in energy at sites 1 and 3, and the $xy$ and $zx$ orbitals are lowered at sites 2 and 4 with large $d$-type Jahn-Teller distortion.

In the case of moderately distorted SmTiO$_3$, the TiO bonds are elongated along the $x$ and $y$ directions while along the $z$-direction, they are contracted in all TiO$_6$ octahedra. As a result, the O crystal field in SmTiO$_3$ energetically lowers the $xy$ orbitals, and consequently favors the $xy$ occupation at every Ti site, while both $yz$ and $zx$ orbitals are higher in energy.

The increased $xy$-occupation is expected to cause a two-dimensional anisotropy in the spin coupling in which the AFM coupling along the $c$-axis is rather weak. Indeed, the calculation of the spin-exchanges showed that when the strength of the O crystal field is comparable to, or is larger than the Sm crystal field, a nearly perfect two-dimensional spin coupling is realized. The observed depression of $T_{\rm N}$ at SmTiO$_3$ may be attributed to this two-dimensional anisotropy. This two-dimensional anisotropy may be observed in SmTiO$_3$.

Now, let us summarize the present section. We have discussed that the $R$ crystal field lifts the $t_{2g}$ degeneracy also in $R$TiO$_3$ where $R=$La, Ce, Pr, Nd and Sm, and stabilizes the AFM(G) ordering as well as the orbital ordering. The spin-exchange constant is gradually decreased with decreasing size of the $R$ ion, which is consistent with the decrease of $T_{\rm N}$ in experiments. Further, with increasing GdFeO$_3$-type distortion, the orbital state in the $R$ crystal field tends to continuously change towards the orbital state in the FM compounds such as GdTiO$_3$ and YTiO$_3$ without changing its symmetry, which is in agreement with the resonant x-ray scattering results. In SmTiO$_3$ with moderate octahedral distortion, the Sm crystal field and the O crystal field compete strongly, which causes a two-dimensional anisotropy in the orbital-spin structure. The depression of $T_{\rm N}$ at $R=$Sm in the phase diagram is attributed to this anisotropy. With decreasing size of the $R$ ion, while the $R$ crystal field dominantly determines the orbital-spin structure at $R=$La, Ce, Pr and Nd, the O crystal field becomes to dominate over the $R$ crystal field at $R=$Gd and Y through the severe competition at $R=$Sm. As observed by several experiments, although the GdFeO$_3$-type distortion is rather large, the orbital structures in GdTiO$_3$ and YTiO$_3$ with strong Jahn-Teller distortion are well characterized by the $d$-type Jahn-Teller distortion, which leads to a FM ground state.

Here, we mention that for quantitative study on the orbital-spin structures in the perovskite titanates, it is necessary to treat the competition and cooperation of the crystal field from the $R$ ions and that from the O ions in more precise way. However, because of the uncertainty due to the dielectric constants $\epsilon_{\rm TiO}$ and $\epsilon_{\rm TiR}$, it is difficult to estimate the contributions of these two crystal fields without ambiguity. This is left for the future studies.

Through the above discussion, it has been clarified that in the perovskite-type transition-metal oxides, the $R$ ions in the GdFeO$_3$-type structure generate the crystal field, which lifts the $3d$ orbital degeneracy. This mechanism of lifting the orbital degeneracy eventually plays an important role in controlling the orbital-spin structures in the perovskite compounds, competing with the usual Jahn-Teller mechanism. This mechanism has a general importance since the GdFeO$_3$-type distortion is a universal phenomenon, which is seen in a large number of perovskite-type compounds. This mechanism may also play important roles on the electronic structures in other perovskite compounds.

\section{Phase diagram and bandwidth-control experiments for the insulating titanates $R$TiO$_3$}

\begin{figure}[tdp]
\caption{Schematic magnetic phase diagram for $R$TiO$_3$ and solid-solution system La$_{1-x}$Y$_x$TiO$_3$ in the plane of temperature and the GdFeO$_3$-type distortion. At $R=$La, Ce, Pr and Nd with relatively small GdFeO$_3$-type distortions, the crystal field (CF) from the $R$ ions dominantly determines the electronic structures, while in significantly distorted GdTiO$_3$ and YTiO$_3$, the O crystal field dominates. In SmTiO$_3$, both crystal fields strongly compete.}
\label{06fig01}
\end{figure}
We are now on the stage to discuss the whole nature of the experimentally obtained magnetic phase diagram~\cite{Greedan85,Katsufuji97,Goral82,Okimoto95} on the basis of the discussions in Secs. 2-4 (see also Fig.~\ref{06fig01}). In this phase diagram, the coupling and interplay of the orbitals and spins are controlled by the GdFeO$_3$-type distortion, resulting in various orbital-spin phases and their phase transition as a function of this distortion.

In LaTiO$_3$ with the smallest GdFeO$_3$-type distortion, since the atomic distortion of the TiO$_6$ octahedron is small, the crystal field from the La ions dominantly lifts the $t_{2g}$ degeneracy and determines the Ti $3d$-orbital state. As a result, LaTiO$_3$ exhibits the AFM(G) ordering with the lowest orbital occupation in the nearly-trigonal La crystal field. The AFM(G) ordering due to the $R$ crystal field is also realized in the compounds of Pr, Ce, Nd and Sm with relatively small octahedral distortion. In addition, as $R$ goes from La to Nd, $T_{\rm N}$ gradually decreases with decreasing spin-exchange constant. The decrease of the spin exchange is caused by the decrease of $t_{2g}$ hybridizations via the O $2p$ states with increasing GdFeO$_3$-type distortion.

In SmTiO$_3$, the TiO$_6$ octahedra are moderately distorted so that the competition between the crystal field from the O ions and that from the Sm ions comes about. In SmTiO$_3$, the O crystal field works as increasing the $xy$-orbital occupation in all Ti sites. As a result, the AFM(G) spin coupling becomes to have a two-dimensional anisotropy. The depression of $T_{\rm N}$ of SmTiO$_3$ may be attributed to this anisotropy.

In GdTiO$_3$ and YTiO$_3$ with strong GdFeO$_3$-type distortion, the electronic structure is dominantly determined by the crystal field from the O ions with the largely distorted TiO$_6$ octahedra. These compounds show the FM ground state with an orbital ordering. This orbital ordering is stabilized both by the $d$-type Jahn-Teller distortion and by the $t_{2g}$-$e_g$ hybridizations induced by the GdFeO$_3$-type distortion. The $t_{2g}$-$e_g$ hybridizations also play an important role on the stability of the FM ordering. We have shown that the FM coupling in the vicinity of the AFM-FM phase boundary has a strong two-dimensional character, resulting in the depressed $T_{\rm C}$. 

Near the AFM-FM phase boundary, both AFM and FM states have a two-dimensional anisotropy, which is expected to depress $T_{\rm N}$ and $T_{\rm C}$. The puzzling second-order like behavior of the AFM-FM transition can be understood from these depressions. In addition, although there is no stoichiometric system between SmTiO$_3$ (AFM(G)) and GdTiO$_3$ (FM), the AFM(A) state might be observed in the solid-solution systems such as La$_{1-x}$Y$_x$TiO$_3$ and Gd$_{1-x}$Sm$_x$TiO$_3$ sandwiched by the AFM(G) and FM phases.

We have pointed out that the GdFeO$_3$-type distortion works as a control mechanism on the orbital-spin structures in the perovskite titanates through introducing the $t_{2g}$-$e_g$ hybridizations or through generating the $R$ crystal field. In addition, we have also pointed out that in the titanates, the $t_{2g}$ degeneracy is inherently lifted by this $R$ crystal field. This allows a single-band description of the low-energy properties of this system as a good starting point. 

Indeed, it has proved that the results of bandwidth-control experiments are well described by the theories based on the single-band Hubbard models. The bandwidth control can be achieved in the end compounds $R$TiO$_3$ or in the solid solutions like La$_{1-x}$Y$_x$TiO$_3$. Since the GdFeO$_3$-type distortion reduces the electron transfers between the neighboring Ti $t_{2g}$ orbitals mediated by the O $2p$ states, we can control the Ti $3d$ bandwidth through this distortion by varying the $R$ ions in $R$TiO$_3$, or in La$_{1-x}$Y$_x$TiO$_3$ by substituting the La ions with Y ions. According to the tight-binding calculation, the bandwidth of the $t_{2g}$ state decreases linearly with increasing magnitude of the GdFeO$_3$-type distortion, or with increasing Y concentration, and the bandwidth for YTiO$_3$ is smaller by about 20 $\%$ relative to that for LaTiO$_3$~\cite{Okimoto95}.

\begin{figure}[tdp]
\caption{Experimentally obtained 2$\Delta_{\rm act}$ for $R$TiO$_3$ is
 plotted as functions of $U/W$. Here, $\Delta_{\rm act}$ is the activation energy of the resistivity, and $U/W$ is the strength of the electron correlation normalized to that of LaTiO$_3$. The solid line is the result of least-square fitting with a linear function. (From Katsufuji, Taguchi and Tokura, Ref.~\cite{Katsufuji97})}
\label{06fig02}
\end{figure}
\begin{figure}[tdp]
\caption{Calculated charge gap as functions of the on-site Coulomb repulsion $U$  (a) for the single-band Hubbard model in the infinite dimension with semicircular density of states (From Georges $et$ $al.$, Ref.~\cite{Georges96}), and (b) for that on the square lattice (From Watanabe and Imada, Ref.~\cite{Watanabe04}). In Fig.(a), the solid line and the dotted line show the result obtained from exact diagonalization and that from iterated perturbation theory. In (b), finite-size charge gaps on $L \times L$ lattices are extrapolated to that in the thermodynamic limit given by the black square with a linear fit. The strength of the Coulomb repulsion $U$ and the charge gap $\Delta$ are normalized by the bandwidth for (a), and by the nearest-neighbor transfer integral for (b).}
\label{06fig03}
\end{figure}
By controlling the bandwidth $W$, the evolution of the Mott-Hubbard gap as a function of the strength of electron correlation ($U/W$) has been investigated~\cite{Crandles92,Katsufuji95,Katsufuji97,Okimoto95}. The magnitudes of the gap were measured by extrapolating the optical conductivity spectra for La$_{1-x}$Y$_x$TiO$_3$~\cite{Okimoto95}. The gaps for $R$TiO$_3$ were also measured from the optical conductivity and the temperature dependence of resistivity~\cite{Katsufuji95,Katsufuji97}. It turned out that the gap $\Delta$ shows a linear increase as a function of $U/W$ as shown in Fig.~\ref{06fig02}. The linear opening of the charge gap has been reproduced in several theoretical studies on the single-band Hubbard models~\cite{Georges96,Watanabe04}. We show the calculated charge gap as a function of the electron interaction $U$ for a single-band Hubbard model in the infinite dimension in Fig.~\ref{06fig03} (a). In addition, this linear behavior has been also reproduced for the single-band Hubbard model on the 2D square lattice as shown in Fig.~\ref{06fig03} (b). These two results imply that the linear growth of the Mott gap is universal irrespective of the spatial dimensionality, if a nonzero transition point $U=U_c$ exists.  The knowledge about the intrinsic lifting of the $t_{2g}$ degeneracy due to the GdFeO$_3$-type distortion in the titanates justifies the correspondence between the experiments and the theories based on the single-band Hubbard models.

\section{Filling-control metal-insulator transition}

As mentioned in Sec.1, the titanate system has been focused as a filling-control metal-insulator transition system. Hole doping to the Ti $3d$ bands is achieved in $R_{1-x}A_x$TiO$_{3+y}$ by substituting the trivalent $R$ ions with the divalent alkaline-earth ions ($A=$Ca, Sr, Ba), or by using the oxygen off-stoichiometry. By finely controlling the hole concentration in the systems such as La$_{1-x}$Sr$_x$TiO$_{3+y}$ and Y$_{1-x}$Ca$_x$TiO$_{3+y}$, the nature of the transition has been intensively investigated, and the critical behavior of the transition and the strongly filling-dependent properties near the transition point have been revealed.

The nature of the filling-control metal-insulator transition has been also theoretically studied. Most of these studies are based on the single-band Hubbard models although in reality, the titanate system with the Ti $3d$ orbitals should be described by a multi-orbital Hubbard model. However, a lot of results obtained by the single-band theories have shown good agreement with the experimental results in spite of their extreme simplifications.

So far, it has been controversial whether the simple single-band description can be justified for this system, and a question why the experimental results are well explained by the theories based on the single-band Hubbard models has not been clearly answered. In the previous section, we have mentioned that in the titanates, the $R$ crystal field due to the GdFeO$_3$-type distortion eventually lifts the $t_{2g}$-orbital degeneracy, which allows the single-band description of the ground-state and low-energy properties as a good starting point. Indeed, as shown in the previous section, the theories based on the single-band Hubbard models well reproduce the linear behavior of the charge gap growth as a function of $U/W$, which was experimentally observed both in the end compounds $R$TiO$_3$ and in the solid solutions La$_{1-x}$Y$_x$TiO$_3$.

In addition to these insulating systems, the inherent lifting of the $t_{2g}$ degeneracy also allows the single-band description of the metallic hole-doped systems in the low-energy region. In this section, we review several filling-control experiments on the hole-doped systems $R_{1-x}A_x$TiO$_3$ by comparing with the corresponding theoretical results. A measurement of the resistivity showed that the metal-insulator transition occurs at $x\sim0.05$ for La$_{1-x}$Sr$_x$TiO$_3$~\cite{Tokura93a,Okada93,Katsufuji94}, at $x\sim0.2$ for Nd$_{1-x}A_x$TiO$_3$~\cite{Eylem95}, and at $x\sim0.4$ for Y$_{1-x}$Ca$_x$TiO$_3$~\cite{Taguchi93,Tokura93b,Katsufuji94}. The Hall coefficient $R_{\rm H}^{-1}$ showed the existence of electron-type carriers (negative $R_{\rm H}^{-1}$), and estimated carrier density is proportional to filling $n=1-x$ implying existence of a large Fermi surface~\cite{Tokura93a,Tokura93b}. Several measurements have shown the strongly filling dependent properties of the hole-doped metallic compounds near the metal-insulator phase boundary in (1) magnetic susceptibility ($\chi$) and specific heat ($C=\gamma T$), (2) resistivity ($\rho$), and (3) optical conductivity ($\sigma(\omega)$). In the following, we discuss the measured results on the above quantities along with the theoretical results each by each.\\

\begin{figure}[tdp]
\caption{Experimentally measured doping dependence of the specific-heat coefficient ($\gamma$) and the magnetic susceptibility ($\chi$) in the metallic region of La$_{1-x}$Sr$_x$TiO$_3$ (lower panel), and the Wilson ratio $\chi/\gamma$ normalized by $3\mu_{\rm B}^2/\pi^2 k_{\rm B}^2$ (upper panel). (From Tokura $et$ $al.$, Ref.~\cite{Tokura93a})}
\label{menhnsE}
\end{figure}
\begin{figure}[tdp]
\caption{Calculated doping dependence of the specific-heat coefficient ($\gamma$) obtained by the dynamical mean-field study on the infinite-dimensional Hubbard model. The experimental result for La$_{1-x}$Sr$_x$TiO$_3$ (Tokura $et$ $al.$, Ref.~\cite{Tokura93a}) is also presented for comparison. (From Rozenberg, Kotliar and Zhang, Ref.~\cite{Rozenberg94}) Upper panel: Calculated Wilson ratio for the infinite-dimensional Hubbard model as a function of doping. (From Georges $et$ $al.$, Ref.~\cite{Georges96})}
\label{menhnsT}
\end{figure}
\noindent
(1) Magnetic susceptibility ($\chi$) and specific heat ($C=\gamma T$)\\
Tokura $et$ $al.$ measured the filling dependence of $\chi$ and $\gamma$ for La$_{1-x}$Sr$_x$TiO$_3$~\cite{Tokura93a,Kumagai93}. Both two quantities are critically enhanced as the system approaches the metal-insulator boundary ($x$=0) from a metallic side with a nearly constant Wilson ratio $R_W=\chi/\gamma$ of $\sim$2 against doping $x$ (see Fig.~\ref{menhnsE}). Similar enhancement was also observed in Y$_{1-x}$Ca$_x$TiO$_3$~\cite{Taguchi93,Tokura93b,Kumagai93} and in Nd$_{1-x}$Ca$_x$TiO$_3$~\cite{Ju94}. These enhancements indicate a critical increase of the effective electron mass. Such a mass renormalization in $R_{1-x}A_x$TiO$_3$ was also observed in several measurements such as resistivity~\cite{Tokura93a}, optical conductivity~\cite{Fujishima92,Taguchi93}, Raman scattering~\cite{Katsufuji94,Katsufuji94b}, photoemission spectra~\cite{Robey95,Yoshida02}, and NMR~\cite{YFurukawa99}. In fact, the mass renormalization near the filling control transition observed here is a generic feature of the filling-control metal-insulator transition, which was predicted theoretically by Furukawa and Imada (Ref.~\cite{Furukawa93}) and Imada (Ref.~\cite{Imada94}). In addition, the experimentally observed enhancement of $\gamma$ is well reproduced by a dynamical mean-field study on the infinite-dimensional Hubbard model~\cite{Georges96,Rozenberg94}. In Fig.~\ref{menhnsT}, the result of Rozenberg $et$ $al.$ (Ref.~\cite{Rozenberg94}) for $\gamma$ is depicted as a function of doping. In this calculation, to consider the Sr-doped La-titanate, the value of $U$ is chosen to be insulating but close to the metal-insulator phase boundary at zero doping. The experimental result of Tokura $et$ $al.$ for La$_{1-x}$Sr$_x$TiO$_3$ (Ref.~\cite{Tokura93a}) is also presented for comparison in the same figure. The figure shows a good agreement between them. 

As for the Wilson ratio, the calculated result shows roughly doping-independent behavior in the intermediate doping region in agreement with the experimental observation. On the other hand, the calculated Wilson ratio vanishes at the transition point while experimentally it remains constant around 2. In addition, for a large doping region, the Wilson ratio becomes the noninteracting value of $\sim$1 in the calculation because the electron correlation is very weak in this region, which is also inconsistent with the experimentally observed constant behavior. These discrepancies may possibly be due to the limitation of the single-band description, and may be solved if we properly take into account the orbital degrees of freedom and the coupling between spins and orbitals.\\

\begin{figure}[tdp]
\caption{Temperature dependence of the resistivity ($\rho$) in metallic La$_{1-x}$Sr$_x$TiO$_3$. (From Tokura $at$ $al.$, Ref.~\cite{Tokura93a})}
\label{resistE}
\end{figure}
\noindent
(2) Resistivity ($\rho$)\\ 
Temperature dependence of the resistivity in La$_{1-x}$Sr$_x$TiO$_3$ for
various filling was measured~\cite{Tokura93a,Okada93,Onoda98}. As shown
in Fig.~\ref{resistE}. the resistivity is characterized by the relation
$\rho=\rho_0+AT^2$, indicating that the system is the Fermi liquid. This
$T^2$ behavior was reproduced in the infinite-dimensional
Hubbard model~\cite{Pruschke93,Jarrel94}. Further, the coefficient $A$
is enhanced as the system approaches the metal-insulator transition
point~\cite{Tokura93a}. It was shown that a ratio $A/\gamma^2$ is
approximately 1.0 $\times$ 10$^{-11}$ $\Omega$cm(molK/mJ)$^2$, which is nearly the same as the universal value for the heavy fermion systems. This value is in fair agreement with the theoretically predicted value in the infinite-dimensional Hubbard model~\cite{Moeller95}.\\

\begin{figure}[tdp]
\caption{Experimentally measured optical conductivity spectra for $R_{1-x}$Sr$_x$TiO$_3$ and $R_{1-x}$Ca$_x$TiO$_3$ with $R=$La, Nd, Sm and Y. Each spectra is labeled by the nominal hole concentration $\delta$. Arrows show the equal-absorption points (see text). (From Katsufuji, Okimoto and Tokura, Ref.~\cite{Katsufuji95})}
\label{optcE1}
\end{figure}
\begin{figure}[tdp]
\caption{The weight of the inner-gap spectra $N_D$ as functions of the hole concentration $\delta$ for $R_{1-x}$Sr$_x$TiO$_3$ and $R_{1-x}$Ca$_x$TiO$_3$ with $R=$La, Nd, Sm and Y. (From Katsufuji, Okimoto and Tokura, Ref.~\cite{Katsufuji95})}
\label{optcE2}
\end{figure}
\noindent
(3) Optical conductivity ($\sigma(\omega)$)\\
The evolution of the Drude part of the optical conductivity as a function of doping also shows agreement between theories and experiments. The Drude weight is defined as the singularity of the conductivity at $\omega=$0 and $T=$0 in the strict sense. However, in reality, the experimental results always suffer from the finite-temperature broadening as well as life-time effects from the impurity scattering. Thus, experimentally the Drude weight is often discussed based on the inner-gap spectral weight of the conductivity. 

The doping-induced changes in the optical conductivity for $R_{1-x}$Sr$_x$TiO$_{3+y}$ and $R_{1-x}$Ca$_x$TiO$_{3+y}$ were investigated~\cite{Katsufuji95,Okimoto95,Katsufuji97}. The optical conductivity for $R=$La, Nd, Sm and Y are displayed in Fig.~\ref{optcE1}. The spectra are labeled with the nominal hole concentration $\delta=x+2y$ where $y$ is the oxygen off-stoichiometry. These figures show that the spectral weights transfer from the Mott-Hubbard-gap region to the inner-gap region with hole doping across an equal-absorption (isosbetic) point around 1.2 eV, leading to the growth of the quasi-Drude weight. The effective number of electrons defined as
\begin{equation}
 N_{\rm eff}(\omega)=\frac{2m_0}{\pi e^2N} \int_0^{\omega} \sigma(\omega') d\omega'
\end{equation}
where $m_0$ is the free electron mass and $N$ is the number of Ti atoms per unit volume, is nearly independent of $\delta$ with $\omega=$2.5 eV, which includes the spectral weight of both the Mott-Hubbard gap excitations and the inner-gap excitations. This implies that the sum of these two parts is nearly conserved irrespective of $\delta$. Therefore, the quantity $N_D=N_{\rm eff}(\omega_c)_{\delta}-N_{\rm eff.}(\omega_c)_{\delta=0}$ with $\omega_c \sim$1.1 eV being the isosbetic point denotes the lower-energy spectral weight transferred from the higher-energy region, or in other words, quasi-Drude weight. In Fig.~\ref{optcE2}, the values of $N_D$ for $R=$La, Nd, Sm and Y are plotted as functions of hole concentration $\delta$, which show the linear increase.

\begin{figure}[tdp]
\caption{Doping dependence of the optical conductivity in the infinite-dimensional Hubbard model with $U=4t$ on the hypercubic lattice with $t/(2\sqrt{d})$ from Jarrell, Freericks and Pruschke (Ref.~\cite{Jarrel95}). The curves correspond to $\delta=$0.068, 0.0928, 0.1358, 0.1878, 0.2455. 0.3, 0.4 and 0.5 from top to bottom curve at high $\omega$. The inset shows the evolution of the Drude weight as a function of doping, which shows a linear increase with doping.}
\label{optcT1}
\end{figure}
\begin{figure}[tdp]
\caption{Total spectral weight of the optical conductivity as a function of the electron concentration $n=1-\delta$ in the infinite-dimensional Hubbard model for the same parameters as in Fig.~\ref{optcT1}. The weight is decomposed into a Drude weight part (open triangles), a mid-infrared part (open squares) and a charge-transfer part (solid triangles). (From Jarrell, Freericks and Pruschke, Ref.~\cite{Jarrel95}.)}
\label{optcT2}
\end{figure}
On the other hand, the optical response was theoretically calculated based on the single-band Hubbard model in the infinite dimension by using the quantum Monte-Carlo method combined with the dynamical mean-field theory~\cite{Jarrel95}. We show in Fig.~\ref{optcT1}, the optical conductivity spectra for small dopings obtained by Jarrell $et$ $al.$ (Ref.~\cite{Jarrel95}). They decomposed the spectral weights into the following three different contributions; (1) Drude part from the transitions within the quasiparticle resonance, (2) mid-infrared part from the transitions from the lower Hubbard band to the unoccupied part of the quasiparticle peak, and (3) charge transfer part from the excitations between the lower and upper Hubbard bands. The weights are presented in Fig.~\ref{optcT2}, where the Drude part linearly increases with increasing hole number $\delta$ (decreasing electron concentration $n=1-\delta$), and the weights associated with the incoherent contributions are concomitantly decreased. The calculation shows a reasonable agreement with the above experimental results of Ref.~\cite{Katsufuji95}.

As shown in this section, the single-band theories have succeeded in describing a number of results of the filling-control experiments. The intrinsic splitting of the $t_{2g}$ levels due to the $R$ crystal field justifies the single-band description and provides a firm ground for the correspondence between the experiments and the single-band theories. We expect that La$_{1-x}$Sr$_x$TiO$_{3+y}$ as well as LaTiO$_3$ are especially well described by the Hubbard model on the cubic lattice where amplitudes of the electron transfers along the $x$-, $y$-, and $z$-directions are nearly the same since in these compounds, the GdFeO$_3$-type distortion is relatively weak, and hence the lowest orbitals are directed nearly along the trigonal directions, which generates nearly isotropic electron transfers between the neighboring lowest orbitals.

However, in YTiO$_3$ and GdTiO$_3$, the lifting of the orbital degeneracy is strongly coupled to the Jahn-Teller distortion in contrast to the case of LaTiO$_3$ where the Jahn-Teller coupling is weak, and is dominated by the $R$ crystal field. In this case, the orbital order may be strongly modified or even lost when the carriers are doped.  Such complexity with contribution from the orbital degrees of freedom combined with dynamical Jahn-Teller fluctuations has to be treated separately beyond the single-band scheme.  In fact, the insulating behavior is extended to a heavily doped region in YTiO$_3$, where the orbital order and Jahn-Teller distortion play crucial roles.

Finally, even for LaTiO$_3$, it may be necessary to take into account the orbital degrees of freedom for higher-energy excitations. For instance, the multi-orbital effect may appear in the higher-energy region of the optical conductivity and the photoemission spectra. Indeed, it was recently claimed that the coherent and incoherent features of the observed photoemission spectra for the doped titanates show disagreement with the results obtained from the single-band Hubbard models~\cite{Fujimori92a,Fujimori92b,Fujimori92c,Robey93,Robey95,Morikawa96,Yoshida02}. To solve the discrepancies, these experimental data should be discussed in the light of theories based on the multi-orbital Hubbard models~\cite{Kotliar96,Kajueter96}. The multi-orbital effects in the higher-energy excitations are issues of interest, and further studies are needed.

\section{Concluding remarks}

The perovskite Ti oxides have been recognized as a key class of materials for understanding the physics of strongly correlated electrons because of the $3d^1$ configuration of the end compounds. The coupling of magnetism and orbitals is one of the most interesting subjects to be clarified in this field, and the titanates have provided an important example for this subject. A rich magnetic and orbital phase diagram has attracted great interest.

In this article, we have reviewed both theoretical and experimental studies and current understanding on the orbital-spin structures in the perovskite titanates as well as their phase transitions. Through these discussions, we have shown that there are several control mechanisms, which determines the orbital-spin structures in the titanates. The obtained knowledges about the orbital-spin couplings and the control mechanisms are expected to be helpful for clarification of electronic structures in other perovskite compounds. For instance, we have discussed that the GdFeO$_3$-type distortion has a universal relevance on the orbital-spin structures through introducing the $t_{2g}$-$e_g$ hybridization (cf. Sec. 3) or through generating the $R$ crystal field (cf. Secs. 4 and 5).  The mechanisms here proposed may work in other perovskite compounds since the GdFeO$_3$-type distortion is a phenomenon which is seen in a large number of perovskite compounds. Thus, we expect that several puzzling properties and phenomena of other perovskite compounds can also be clarified in the future research if we consider these mechanisms. The knowledges may be important also from the aspect of material design. Manipulation of the coupling of spin, orbital and lattice degrees of freedom through these control mechanisms may open a possibility of controlling the dimensionality, transport properties and cooperative response to the external fields of these materials. 

In addition to the coupling of magnetism and orbitals, the metal-insulator transition (Mott transition) is another important subject of the physics of strongly correlated electrons. There are two different routes to achieve the Mott transition. One is the bandwidth control (control of the electron correlation strength $U/t$) and the other is the filling control. As we have discussed in this article, the titanate system provides typical examples of these issues as well. The filling-control Mott transition in the hole-doped titanates has been intensively studied. On the other hand, the bandwidth control can also be achieved in the insulating titanates, and the charge gap evolution as a function of bandwidth has been investigated although the bandwidth cannot be sufficiently large to make the system metallic.

We have discussed the results of bandwidth-control experiments and those of filling-control experiments in the light of the theoretical results based on the single-band Hubbard models. The single-band description of this system as a starting point is justified as far as for the ground-state and the low-energy properties since the Ti $t_{2g}$-orbital degeneracy is inherently lifted by the $R$ crystal field in the GdFeO$_3$-type structure. It has turned out that the titanate system embodies the essential physics of the single-band Hubbard model which is one of the simplest and the most important model capturing the competition of the electron interactions and the kinetic energies in the strongly correlated electron systems.

As discussed in this article, the intensive studies have elucidated the ground-state electronic structures of the perovskite titanates. On the other hand, clarification of the properties of excited states is one of the important subject of further works. For instance, the photoemission spectroscopy and the optical conductivity measurement are expected to detect the higher-energy excitations. To analyze the spectra obtained in the experiments, it may be necessary to take into account the multi-orbital effects on the excitations. Indeed, the filling dependence of photoemission spectra shows a discrepancy between the experiments and the single-band theories as we have mentioned in Sec. 7. It is interesting to study the spectra based on the multi-orbital Hubbard model. In addition, the possibility of detecting the collective orbital excitation (so-called ``orbital wave'') in the titanates has recently been proposed theoretically~\cite{Ishihara04}. Prediction of the nature of orbital-wave spectrum or its experimental detection are also interesting. For the future research on such properties and phenomena associated with the excitations, the understanding of the ground-state electronic structures discussed in this article will be essentially important.

\begin{figure}[tdp]
\includegraphics[scale=1.0]{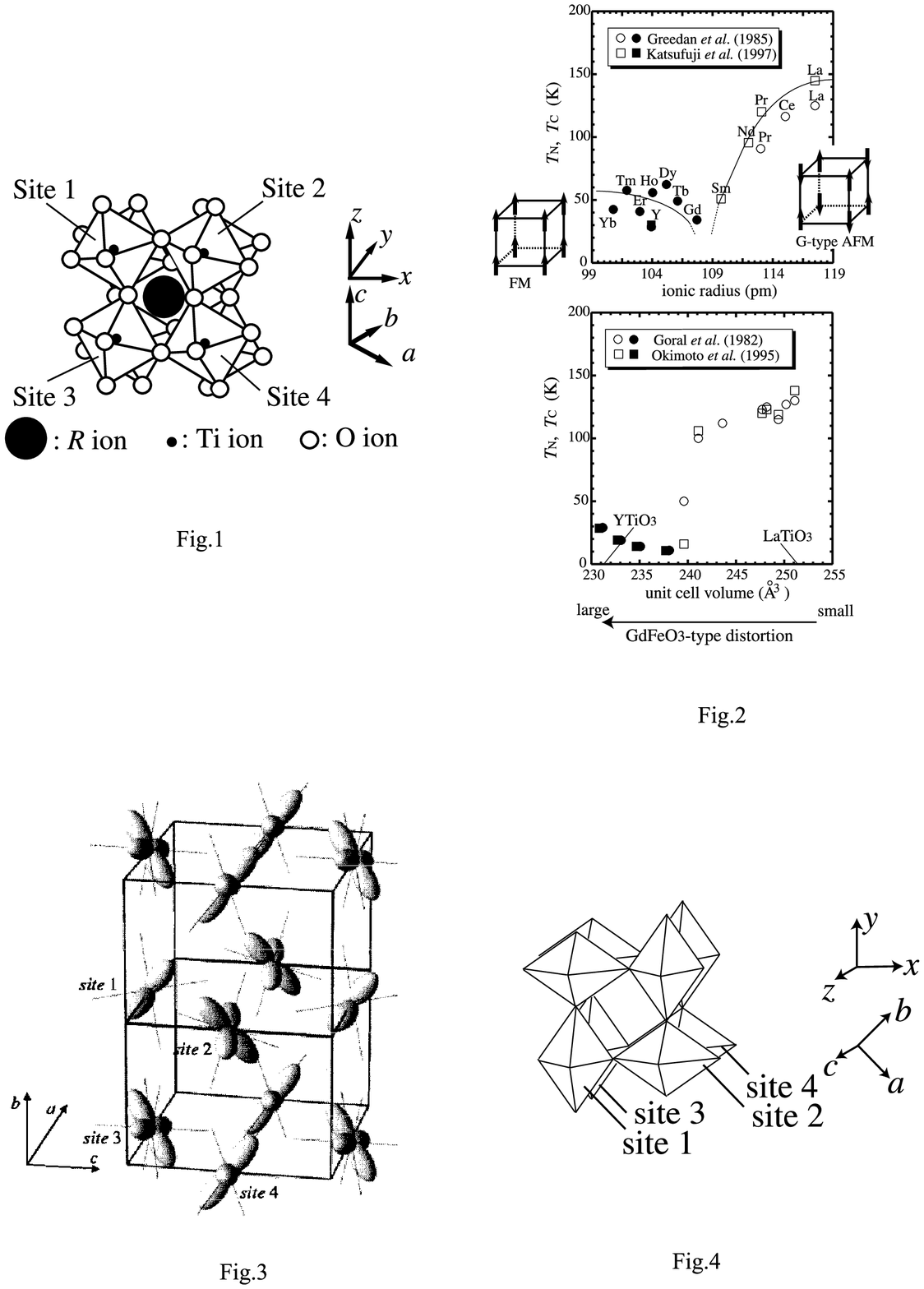}
\end{figure}
\begin{figure}[tdp]
\includegraphics[scale=1.0]{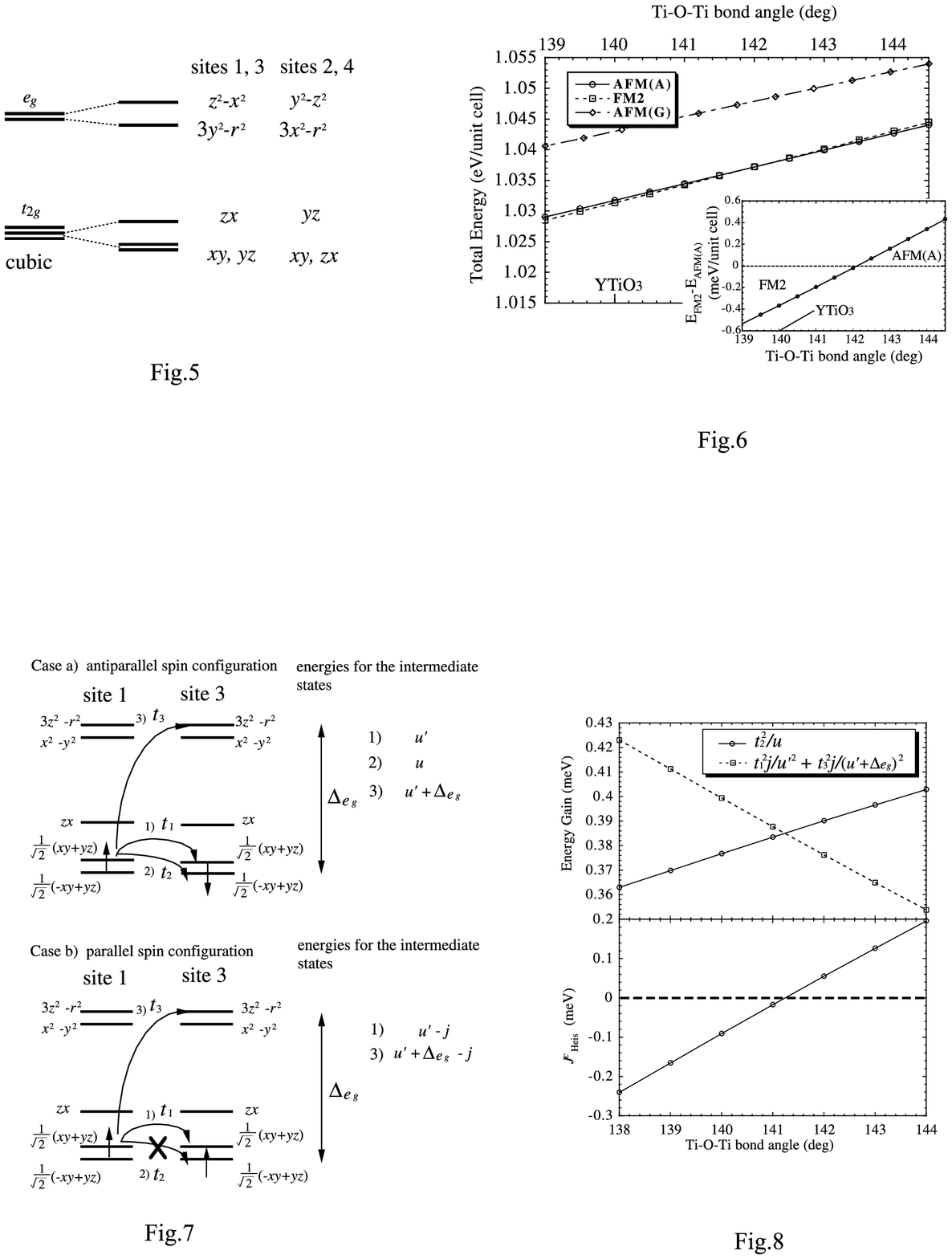}
\end{figure}
\begin{figure}[tdp]
\includegraphics[scale=1.0]{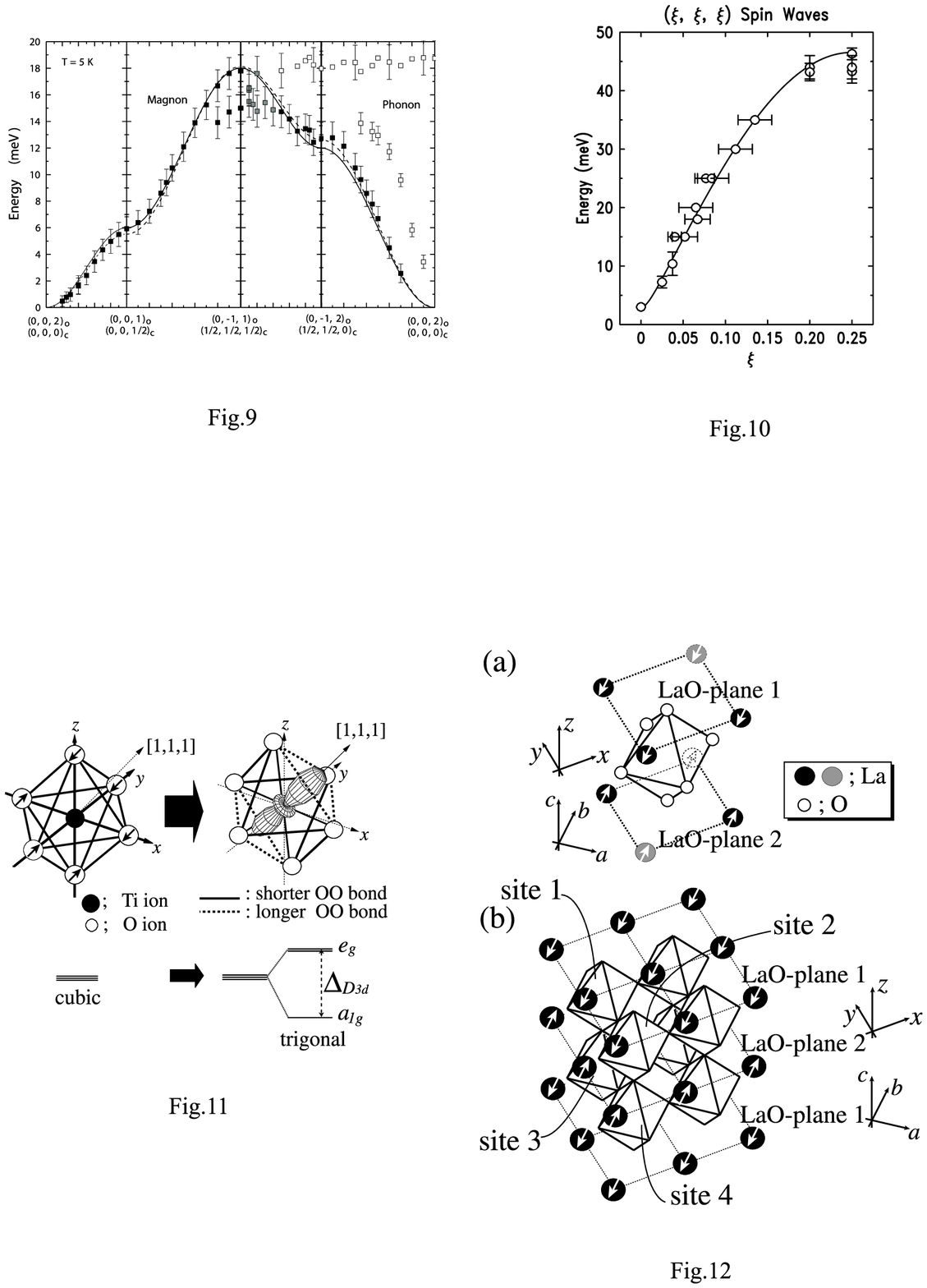}
\end{figure}
\begin{figure}[tdp]
\includegraphics[scale=1.0]{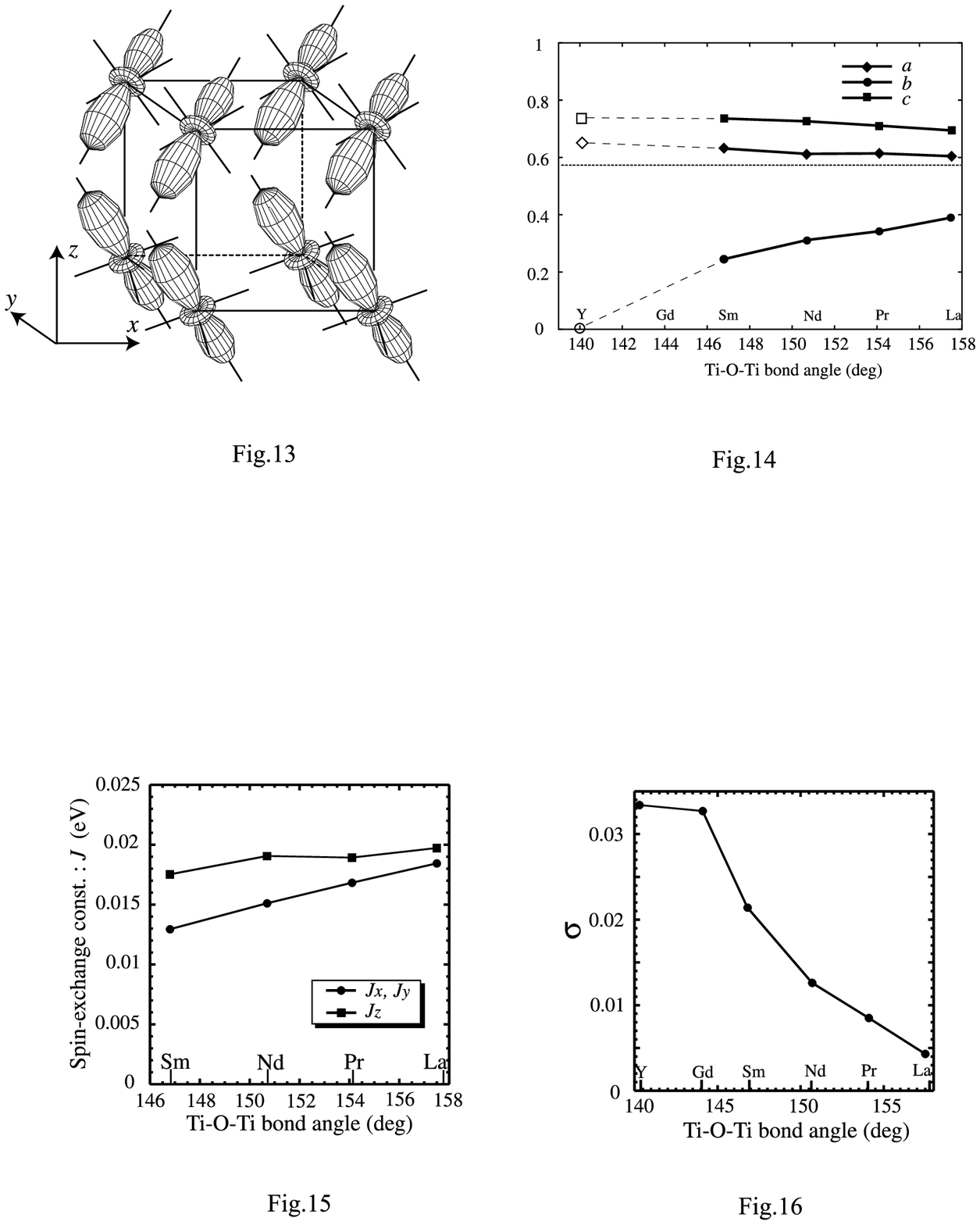}
\end{figure}
\begin{figure}[tdp]
\includegraphics[scale=1.0]{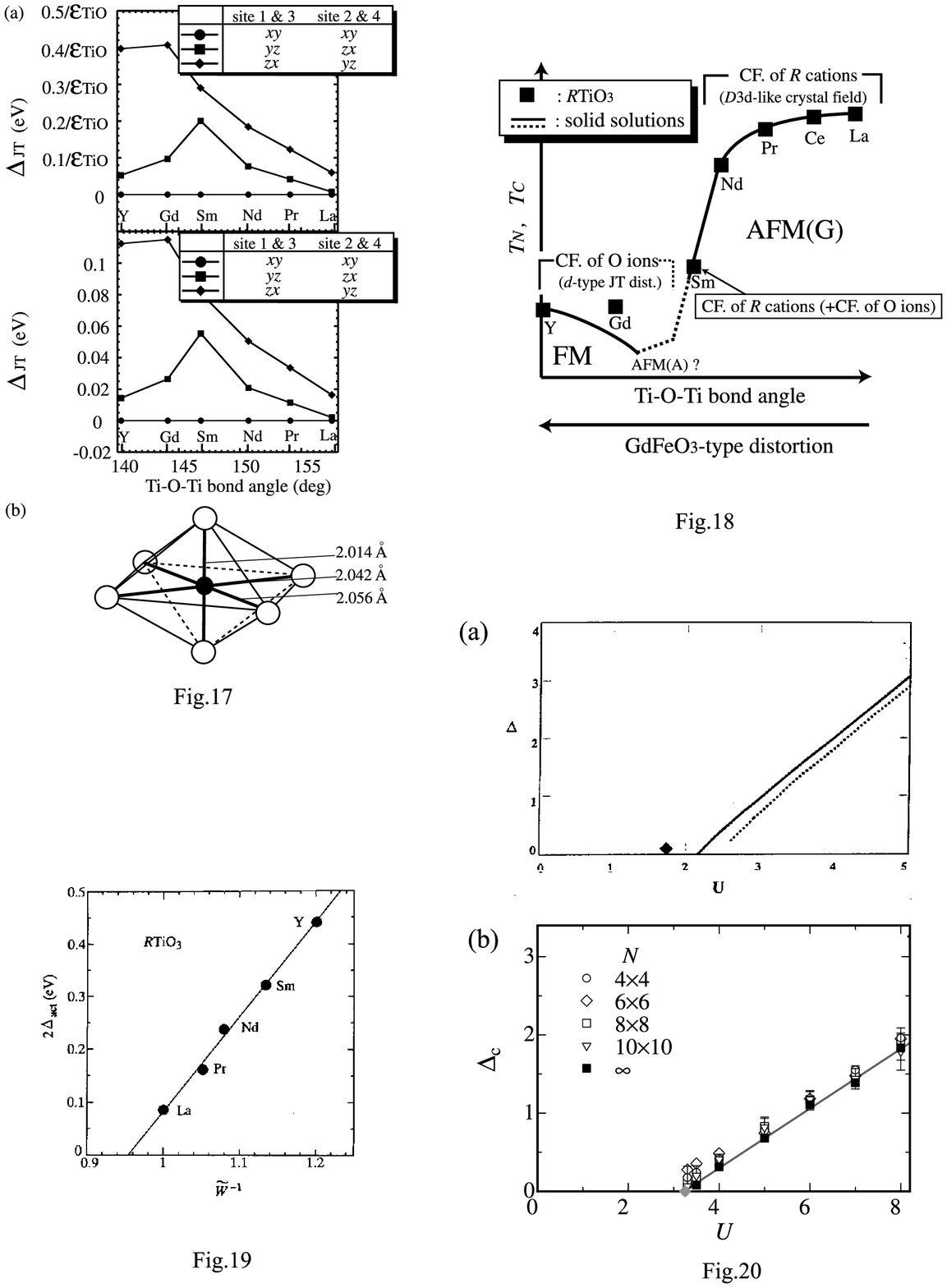}
\end{figure}
\begin{figure}[tdp]
\includegraphics[scale=1.0]{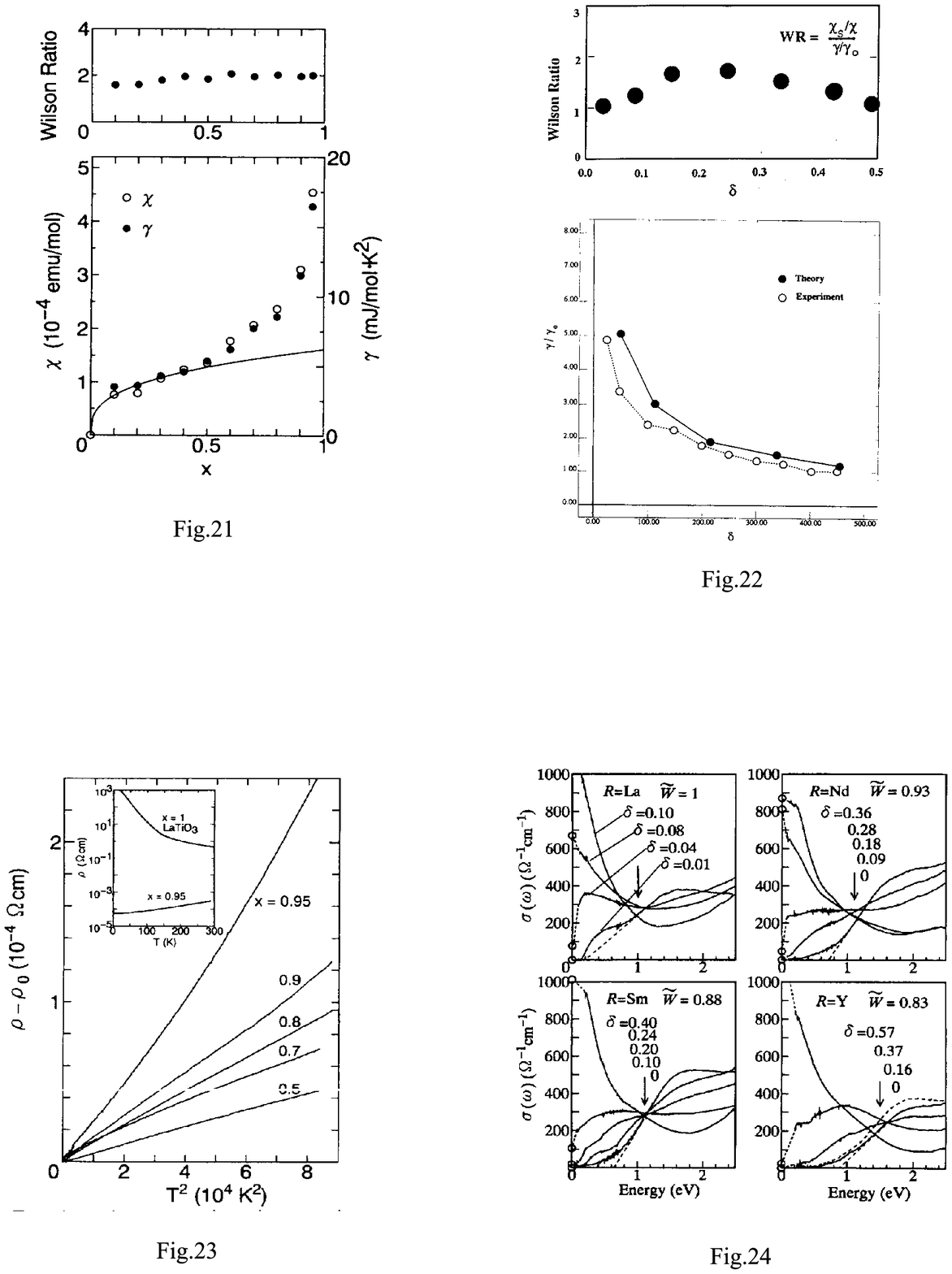}
\end{figure}
\begin{figure}[tdp]
\includegraphics[scale=1.0]{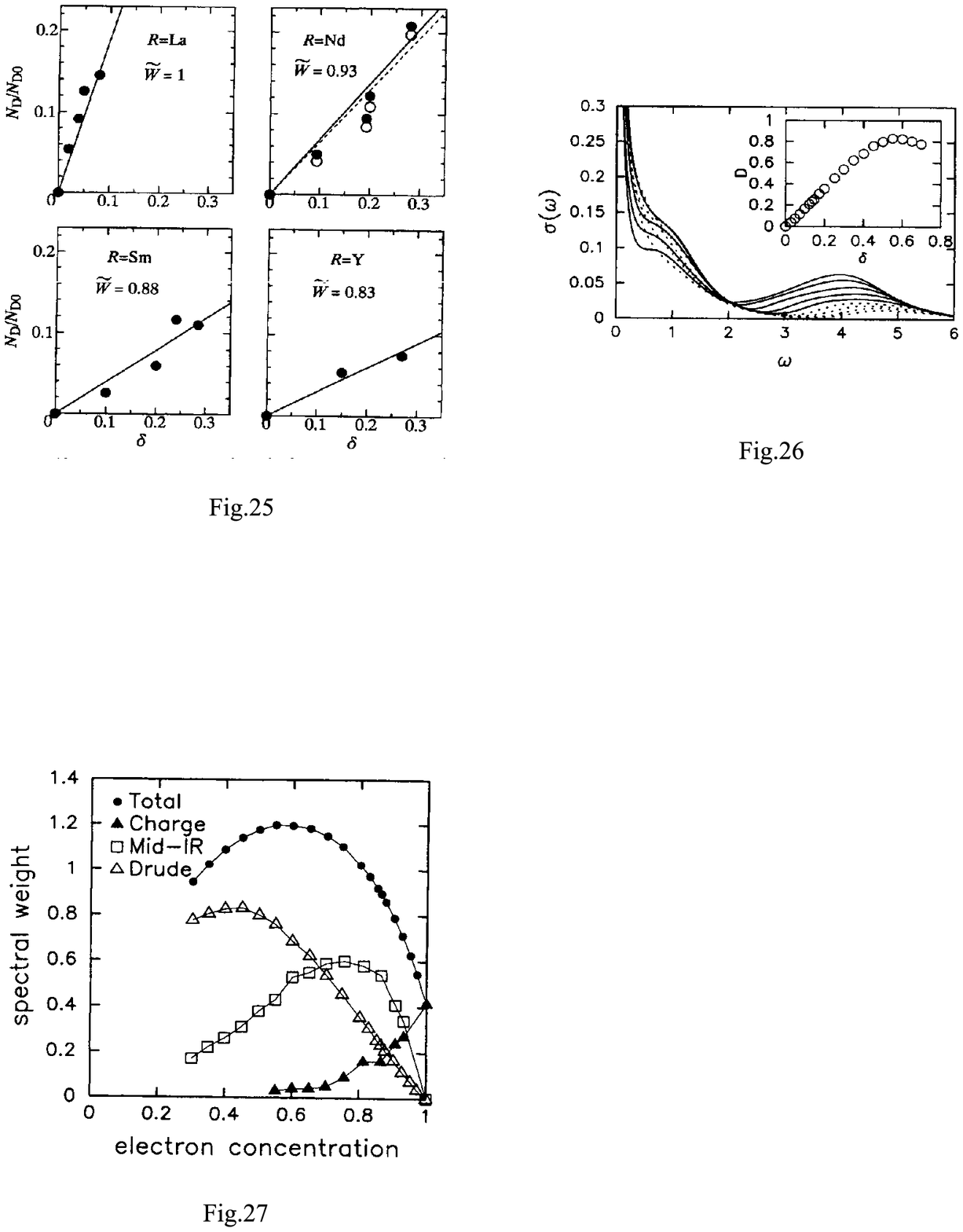}
\end{figure}
\end{document}